\documentclass[11pt]{article}
\usepackage {amsfonts}
\usepackage {graphicx}
\usepackage {epsfig}
\usepackage{amssymb}
\usepackage{amsmath}
\usepackage {longtable}
\usepackage[english]{babel}
\begin{document}

\title{Birefringence effect in the nuclear pseudoelectric field of matter and
an external electric field for a deuteron (nucleus) rotating in a
storage ring}

\author{V.G. Baryshevsky, \\ Research Institute for Nuclear Problems, Belarusian State
University,\\ 11 Bobruyskaya Str., Minsk 220050, Belarus,
\\ e-mail: bar@inp.minsk.by}

\maketitle

\begin{center}
\begin{abstract}
The birefringence effect in the nuclear pseudoelectric field of
matter and an external electric field for a particle (deuteron,
nucleus) moving in a storage ring is discussed. The influence of
the birefringence effect on the EDM measurement experiments is
considered. The attention is drawn to the possibility to measure
the spin-dependent amplitude of the elastic coherent scattering of
a deuteron by a nucleus, the electric polarizability of a deuteron
(nucleus). Using a gas target with polarized nuclei also allows to
study P-,T-odd interactions.
\end{abstract}
\end{center}
%\maketitle

 \section{INTRODUCTION}

The phenomena of spin rotation and spin dichroism (birefringence
effect) for particles with the spin $S\geq1$ in an unpolarized
medium were theoretically described for the first time in
\cite{1,2}.
%--------------------- add -----------------------------------------------------------------------------
%
The peculiarity of this phenomenon is in conversion of the vector
polarization to the tensor one and vice versa (similar  the linear
to circular polarization conversion for a photon in an optically
anisotropic medium - the well known optical birefringence).
However, unlike the optical birefringence, the birefringence
effect for particles appear in isotropic matter
 (and even the spin of matter nuclei is either zero or unpolarized !). Anisotropy is
 provided by the particle itself
 (a particle with the spin $S \ge 1$ and mass $M \ne 0$ has the intrinsic anisotropy).
 Deuteron spin dichroism was recently observed with the $20$ MeV
accelerator \cite{3}.
 Further investigations of this phenomenon are planned to be carry
out with a storage ring and an external beam \cite{4,5}.
{Observation of particle spin rotation and spin dichroism} (the
birefringence effect) with a storage ring requires reducing of
$(g-2)$ precession frequency ($g$ is the gyromagnetic ratio).
 This precession appears due to interaction of the particle magnetic moment
 with an external electromagnetic field.
 The requirement for $(g-2)$ precession cancellation also arises when searching for a deuteron electric
dipole moment (EDM)  by the deuteron spin precession in an
electric field in a storage ring\cite{6, project}.
%
% ---------- addition 1
Some extremely interesting ideas providing noticeable cancellation
for $(g-2)$ precession are proposed in \cite{6,project}

 According to \cite{6} balancing the energy of the
particle and the strength of the electric field in the storage
ring provides to reduce and even zeroize the $(g-2)$ precession
frequency.
As a result, EDM-caused spin rotation grows linearly with time
\cite{6,project}.
 Note that, when $(g-2)$ precession is suppressed, the angle of spin rotation induced by the birefringence effect
 grows linearly with time, too.

 The effect of deuteron (nucleus) birefringence in matter reveals itself in a
storage ring due to presence of the residual gas inside the
storage ring and use of a gas jet (gas target) for deuteron
(nucleus) polarization analysis.
 Moreover, birefringence also occurs in the
solid target, which is used for analysis of polarization of the
deuteron (nucleus) beam outside the storage ring.
Therefore, the phenomenon of birefringence in the gas medium and
polarimeter would appear as a systematic error in the EDM
measurements \cite{project}.
% Therefore, the phenomenon of birefringence in the gas medium and polarimeter
%could hamper the extraction of the EDM-induced signal from the
%background caused by the birefringence effect, when searching the
%EDM.
%
 In addition, study of the birefringence phenomenon is of self-importance
since it makes possible to measure the spin-depended part of the
forward scattering amplitude.

Lastly, the action of the electric field on the deuteron rouses
one more mechanism of deuteron spin rotation and oscillations (the
phenomenon of birefringence in an electric field)
 conditioned by the deuteron
tensor electric polarizability \cite{7}.

In this paper  several our preceding works concerning the
birefringence effect in the nuclear pseudoelectric field of matter
and an external electric field \cite{1,2,3,4,5,7,nastya,kostya}
are summarized for a particle (deuteron, nucleus) moving in a
storage ring . The attention is drawn to the additional
possibility to measure the spin-dependent amplitude of the elastic
coherent scattering of a deuteron by a nucleus, the electric
polarizability of a deuteron (nucleus). Use of a gas target with
polarized nuclei also allows to study P-,T-odd interactions.

%----------- section 2 ------------------ from the file WIND-PhRev_v2.tex  ---------------------literature correction
\section{The index of refraction and effective potential energy of particles in medium.}

Close connection between the coherent elastic scattering amplitude
at zero angle $f(0)$ and the refraction index of medium $N$ has
been established as a result of numerous studies (see, for
example, \cite{12, Goldberger}):
\begin{equation}
N=1+\frac{2\pi \rho }{k^{2}}f\left( 0\right) \label{refr_ind}
\end{equation}
where $\rho $ is the number of particles per $cm^{3}$, $k$ is the
wave number of a particle incident on a target.

The expression (\ref{refr_ind}) was derived in assumption that
$N-1 \ll 1$. If $k \rightarrow 0$ then $(N-1)$ grows and
expression for $N$ has the form
\[
N^2=1+\frac{4\pi \rho }{k^{2}}f\left( 0\right) \label{refr_ind2}
\]

Let us consider particle refraction on the vacuum-medium boundary.

The wave number of the particle in vacuum is denoted $k$,
$k^{\prime} = k N$ is the wave number of the particle in the
medium.

\begin{figure}[htbp]
\epsfxsize = 5 cm \centerline{\epsfbox{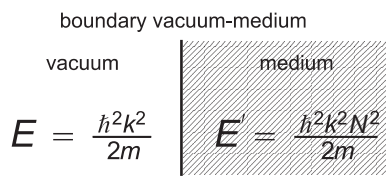}}
\caption{Kinetic energy of a particle in vacuum is not equal to
that in medium.}
\end{figure}

As it can be seen, the kinetic energy of a particle in vacuum $E =
\frac{\hbar^2 k^2}{2 m}$ is not equal to that in the medium
$E^{\prime} = \frac{\hbar^2 k^2 N^2}{2 m}$.

From the energy conservation condition we immediately obtain the
necessity to suppose that the particle in the medium possesses an
effective potential energy (see the detailed theory in
\cite{Goldberger}.) This energy can be easily found  from the
evident equality
\[
E=E^{\prime}+V
\]
i.e.
\begin{equation}
V=E-E^{\prime }=- \frac{2 \pi {\hbar}^2}{m} {\rho} f(0).
\label{U1}
\end{equation}

Above we considered the rest target. But in storage rings moving
bunches can be used as a target. Therefore we should generalize
the expressions (\ref{refr_ind},\ref{U1}) for this case. Thus, let
us consider the collision of two bunches of particles. Suppose
that in the rest frame of the storage ring the particles of the
first beam have the energy $E_1$ and Lorentz-factor $\gamma_1$,
whereas particles of the second beam are characterized by the
energy $E_2$ and Lorentz-factor $\gamma_2$. Let us recollect that
the phase of a wave in a medium is Lorentz-invariant. Therefore,
we can find it by the following way. Let us choose the reference
frame, where the second beam rests. As in this frame particles of
the second beam rest, then the refraction index can be expressed
in the conventional form (\ref{refr_ind}):
\begin{equation}
N_1^{\prime }=1+\frac{2\pi \rho_2^{\prime } }{{k_1^{\prime
}}^{2}}f\left(E_1^{\prime}, 0\right), \label{refr_ind3}
\end{equation}
where $\rho_2^{\prime }=\gamma_2^{-1} \rho_2$ is the density of
the bunch 2 in its rest frame and $\rho_2$ is the density of the
second bunch in the storage ring frame, $k_1^{\prime}$ is the
wavenumber of particles of the first bunch in the rest frame of
the bunch 2. Let us denote the length of the bunch 2 in its rest
frame as $L$, $L=\gamma_2 ~ l$, where $l$ is the length of this
bunch in the storage ring frame.

Now we can find the change of the phase of the wave caused by the
interaction of the particle 1 with the particles of bunch 2:
\begin{equation}
\phi=k_1^{\prime}(N_1^{\prime}-1)L=\frac{2 \pi
\rho_2^{\prime}}{k_1^{\prime}} f(E_1^{\prime},0)~L =\frac{2 \pi
\rho_2}{k_1^{\prime}}{f(E_1^{\prime},0)}{k_1^{\prime}}~l  ,
\label{phase}
\end{equation}

It is known that the ratio
$\frac{f(E_1^{\prime},0)}{k_1^{\prime}}$ is invariant, so we can
write
$\frac{f(E_1^{\prime},0)}{k_1^{\prime}}=\frac{f(E_1,0)}{k_1}$,
where $f(E,0)$ is the amplitude of elastic coherent forward
scattering of particle 1 by the moving particle 2 in the rest
frame of the storage ring.

As a result
\begin{equation}
\phi=\frac{2 \pi \rho_2}{k_1} f(E_1,0)~l=\frac{2 \pi \rho_2}{k_1}
f(E_1,0)~v_{rel}~t, \label{phase1}
\end{equation}
where $v_{rel}$ is the velocity of relative motion of the particle
1 and bunch 2, if the first particle is nonrelativistic, whereas
the second one is relativistic, then $v_{rel}\approx c$, where $c$
is the speed of light, $t$ is the time of interaction of the
particle 1 with the bunch 2 in the rest frame of the storage ring.

The particle with the velocity $v_1=\frac{\hbar k_1 c^2}{E_1}$
passes the distance $z=v_1~t$ over the time $t$. It should be
noted that the path length $z$ differs from the length of bunch 2,
because it moves. Expression (\ref{phase1}) can be rewritten as:
\begin{equation}
\phi=\frac{2 \pi \rho_2}{k_1} f(E_1,0)~\frac{v_{rel}}{v_1}~z=k_1
(N_1-1)z, \label{phase_z}
\end{equation}
where the index of refraction of the particle 1 by the beam of
moving particles 2 is:
\begin{equation}
N_1=1+\frac{2 \pi {\rho}_2}{{k_1}^2}~\frac{v_{rel}}{v_1}f(E,0)
\label{r_ind_z}
\end{equation}
If $v_2=0$, the the conventional expression (1) follows from
(\ref{r_ind_z})

So, let us consider particles captured to a trap. They are
nonrelativistic. From (\ref{U1}) the effective potential energy
$V$ can be expressed as:
\begin{eqnarray}
V=- \frac{2 \pi {\hbar}^2}{m_1} {{\rho}_2} \frac{c}{v_1}f(E_1,0)=
\\ \nonumber =-2 \pi {\hbar} {\rho}_2 c  \frac{f(E_1,0)}{k_1}=\\
\nonumber =-2 \pi {\hbar} {\rho}_2 c
\frac{f(E_1^{\prime},0)}{k_1^{\prime}}=\\ \nonumber =-\frac{2 \pi
{\hbar}^2}{m_1 \gamma_2} {\rho}_2 f(E_1^\prime,0). \label{U_}
\end{eqnarray}
where $\gamma_2$ is the Lorentz-factor of the bunch 2 and the
Lorentz-factor of the particle 1 is $\gamma_1=1$ $\hbar
k_1^{\prime}=m_1 {c} \gamma_2$.

It should be noted that the amplitude of coherent elastic
scattering at zero angle depends on the T-matrix as follows:
\begin{equation}
f(E_1^{\prime},0)= - \frac{(2 \pi)^2 \hbar
k_1^{\prime}}{v_1+v_2}T(E_1^\prime)= (2 \pi)^2 m_1 \gamma_2
T(E_1^\prime) \label{T-matrix}
\end{equation}
i.e. the amplitude of forward scattering  is proportional  to the
Lorentz-factor $\gamma_2$. As a result, the quantity
$\gamma_2^{-1} f(E_1^{\prime},0)$ depends on {the energy of the
particle} only due to possible dependence of T-matrix on energy.
From (\ref{U_},\ref{T-matrix}) one can obtain
\begin{equation}
V=(2 \pi)^3 \rho_2 T(E_1^\prime) \label{T-matrix}
\end{equation}

% ---------------------------------------end from WIND

%--------------------------------------------from file Cosy experiment for PR 2005-03-31.tex --- literature correction
\section{THE PHENOMENON OF ROTATION AND OSCILLATION OF DEUTERON (NUCLEAR) SPIN IN UNPOLARIZED
MATTER (BIREFRINGENCE AND SPIN DICHROISM)}

Let us consider the refraction of a particle with the spin $S \ge
1$ in matter.

 According to \cite{1,2} the index of refraction for such a
particle depends on the particle spin and can be written as:

\begin{equation}
\label{hatn1} \hat {N} = 1 + \frac{{2\pi \rho} }{{k^{2}}}\hat
{f}\left( {0} \right),
\end{equation}
where$\hat {f}\left( {0} \right) = Tr\hat {\rho} _{J} \hat
{F}\left( {0} \right)$, $\rho $~is the density of matter (the
number of scatterers in 1 cm$^3$), $k$~is the deuteron wave
number, $\hat {\rho} _{J} $~is the spin density matrix of the
scatterers, $\hat {F}\left( {0} \right)$~is the operator of the
forward scattering amplitude, acting in the combined spin space of
the particle and scatterer spin $\vec {J}$.

For a particle with the spin $S=1$ (for example, deuteron) in an
unpolarized target $ \hat {f}(0)$ can be written as:
\begin{equation}
\label{hatf} \hat {f}\left( {0} \right) = d + d_{1} (\vec{S}
\vec{n})^{2}.
\end{equation}
where $\vec{S}$ is the deuteron spin operator, $\vec{n}$ is the
unit vector along the deuteron momentum $\vec{k}$.

The axis of quantization \textit{z} is directed along
%$\vec {n} =\frac{{\vec {k}}}{{k}}$, where \textit{}
the particle wave vector $\vec {k}$.
Considering only strong interactions, which are invariant to the
parity transformation and time reversal, we may omit the terms
containing $S$ in odd degrees.
Therefore, the refractive index for deuterons
\begin{equation}
\label{hatn2}
 \hat {N} = 1 +
\frac{{2\pi \rho} }{{k^{2}}}\left( {d + d_{1} S_{z}^2}  \right)
\end{equation}
depends on the deuteron spin orientation relative to the deuteron
momentum.

The refractive index for a particle in the state, which is the
eigenstate of the operator ${S}_{z}$ of spin projection on the
axis $z$:
\begin{equation}
\label{hatn3} \hat {N} = 1 + \frac{{2\pi \rho} }{{k^{2}}}\left( {d
+ d_{1} m_{}^{2} }  \right),
\end{equation}
$m$ is the magnetic quantum number.

\begin{figure}[!h]
\epsfxsize = 5 cm \centerline{\epsfbox{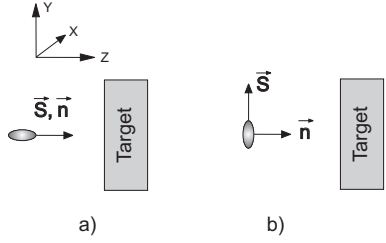}} \caption{Two
possible orientation of vectors $\vec {S}$ and $\vec
{n}=\frac{\vec{k}}{k}$: a) $\vec {S} {\|} \vec {n}$; b) $\vec {S}
{\bot} \vec {n}$}
 \label{sigma}
\end{figure}

According to Eq.(\ref{hatn3}), the refractive indices for the
states with $m=+1$ and $m=-1$ are the same, while those for $m=\pm
1$ and $m=0$ are different ($\Re{\textit{N}(\pm 1)} \neq
\Re{\textit{N}(0)}$ and $\Im{\textit{N}(\pm 1)} \neq
\Im{\textit{N}(0)}$).
This can be obviously explained as follows (see Fig.\ref{sigma}):
the shape of a deuteron in the ground state is non--spherical.
Therefore, the scattering cross-section $\sigma_{\|}$ for a
deuteron with $m= \pm 1$ (deuteron spin is parallel to its
momentum $\vec{k}$) differs from the scattering cross-section
$\sigma_\bot$ for a deuteron with $m=0$:
\begin{equation}
\sigma_{\|} \ne \sigma_\bot ~\Rightarrow ~
 \label{eq7} \Im\textit{f}_\|(0)=\frac{k}{4\pi}
\sigma_{\|}\neq \Im\textit{f}_ \bot (0)=\frac{k}{4\pi}\sigma_\bot.
\end{equation}
According to the dispersion relation $\Re{\textit{f}(0)} \sim\Phi
(\Im{\textit{f}(0))}$, then $\Re{\textit{f}_{ \bot }(0)} \neq
\Re{\textit{f}_{\|}(0)}$.

From the above it follows that deuteron spin dichroism appears
even when the deuteron passes through an unpolarized target: due
to different absorption, the initially unpolarized beam acquires
polarization or, yet more precisely, alignment.

%%%%%%%%%%%%%%%%%%%%%%%

Let us consider the behavior of the deuteron spin in a target.
The spin state of the deuteron is described by its vector and
tensor
 polarization ($\vec {p} = \langle \vec {S}\rangle$ and $p_{ik} =
 \langle Q_{ik} \rangle$,
respectively, here $\hat {Q}_{ij} = \frac{{3}}{{2}}\left( {\hat
{S}_{i} \hat {S}_{j} + \hat {S}_{j} \hat {S}_{i} -
\frac{{4}}{{3}}\delta _{ij}} \right)$.
When the deuteron moves in matter its vector and tensor
polarization appears changed.
To calculate $\vec {p}$ and $p_{ik}$ one need to know the explicit
form of the deuteron spin wave function $\psi$

The wave function of the deuteron that has passed the distance $z$
inside the target is:
\begin{equation}
\label{psiz} \psi \left( {z} \right) = \exp\left( {ik\hat{N}z}
\right)\psi _{0},
\end{equation}
where $\psi_{0}$ is the wave function of the deuteron before
entering the target.
The wave function $\psi$ can be expressed as a superposition of
the basic spin functions $\chi_{m}$, which are the eigenfunctions
of the operators $\hat{S}^{2}$ and $\hat{S}_{z}$ ($\hat {S}_{z}
\chi _{m} = m\chi _{m}$):
\begin{equation}
\label{psi} \psi = \sum\limits_{m = \pm 1,0} {a^{m}\chi _{m}}  .
\end{equation}
Therefore,
\begin{equation}
 \label{psidepth}
\begin{array}{l}
 \Psi = \left( \begin{array}{*{20}c}
 {a^{1}} \hfill \\
 {a^{0}} \hfill \\
 {a^{ - 1}} \hfill \\
\end{array}  \right) = \left( {{\begin{array}{*{20}c}
 {ae^{i\delta _{1}} e^{ikN_{1} z}} \hfill \\
 {be^{i\delta _{0}} e^{ikN_{0} z}} \hfill \\
 {ce^{i\delta _{ - 1}} e^{ikN_{ - 1} z}} \hfill \\
\end{array}} } \right) = \\ \\ \qquad {}\left( {{\begin{array}{*{20}c}
 {ae^{i\delta _{1}} e^{ikN_{1} z}} \hfill \\
 {be^{i\delta _{0}} e^{ikN_{0} z}} \hfill \\
 {ce^{i\delta _{ - 1}} e^{ikN_{1} z}} \hfill \\
\end{array}}} \right),
\end{array}
\end{equation}
(according to the above $N_{1}=N_{-1}$).

Suppose plane $(yz)$ coincides with the plane formed by the
initial deuteron vector polarization $\vec {p}_0 \neq 0$ and the
momentum $\vec{k}$ of the deuteron. In this case $\delta
_{1}-\delta _{0}= \delta _{0}-\delta _{-1}=\frac{{\pi}}{{2}}$, and
the components of the polarization vector at $z = 0$ are $p_x =
0,p_y\neq 0, \mbox{and }p_z\neq 0$.

The components of the vector polarization $\vec{p}=\langle \vec
{S}\rangle = \frac{{\langle \Psi \left| {\vec {S}} \right|\Psi
\rangle} }{{\left\langle {{\Psi} } \mathrel{\left| {\vphantom
{{\Psi} {\Psi} }} \right. \kern-\nulldelimiterspace} {{\Psi} }
\right\rangle} }$ inside the target are:
\\
\begin{eqnarray}
p_x&=& \frac{{ \sqrt {2} e^{ - \frac{{1}}{{2}}\rho \left( {\sigma
_{0} + \sigma _{1}} \right)z}b\left( {a - c} \right)\sin\left(
{\frac{{2\pi \rho} }{{k}}\Re d_{1} z} \right)}}{{\left\langle
{{\Psi} } \mathrel{\left| {\vphantom {{\Psi}  {\Psi} }} \right.
\kern-\nulldelimiterspace}
{{\Psi} } \right\rangle} } ,\nonumber\\
p_y&=&\frac{{\sqrt {2} e^{ - \frac{{1}}{{2}}\rho \left( {\sigma
_{0} + \sigma _{1}} \right)z}b\left( {a + c} \right)\cos\left(
{\frac{{2\pi \rho} }{{k}}\Re d_{1} z} \right)}}{{\left\langle
{{\Psi} } \mathrel{\left| {\vphantom {{\Psi}  {\Psi} }} \right.
\kern-\nulldelimiterspace}
{{\Psi} } \right\rangle} } ,\nonumber \\
p_z& =&\frac{{e^{\rho \sigma _{1} z}\left( {a^{2} - c^{2}}
\right)}}{{\left\langle {{\Psi} } \mathrel{\left| {\vphantom
{{\Psi}  {\Psi }}} \right. \kern-\nulldelimiterspace} {{\Psi} }
\right\rangle} }.\label{rot1}
\\ \nonumber
\end{eqnarray}
Similarly, the components of the tensor polarization $\hat
{Q}_{ij} = \frac{{3}}{{2}}\left( {\hat {S}_{i} \hat {S}_{j} + \hat
{S}_{j} \hat {S}_{i} - \frac{{4}}{{3}}\delta _{ij}} \right)$ are
expressed as:
\\
\begin{eqnarray}
p_{xx}&=&\frac{{ - \frac{{1}}{{2}}\left( {a^{2} + c^{2}}
\right)e^{ - \rho \sigma _{1} z} + b^{2}e^{ - \rho \sigma _{0} z}
- 3ace^{ - \rho \sigma _{1} z}}}{{\left\langle {{\Psi} }
\mathrel{\left| {\vphantom {{\Psi}  {\Psi} }} \right.
\kern-\nulldelimiterspace} {{\Psi} } \right\rangle} }  ,\nonumber
\\
p_{yy}&=&\frac{{ - \frac{{1}}{{2}}\left( {a^{2} + c^{2}}
\right)e^{ - \rho \sigma _{1} z} + b^{2}e^{ - \rho \sigma _{0} z}
+ 3ace^{ - \rho \sigma _{1} z}}}{{\left\langle {{\Psi} }
\mathrel{\left| {\vphantom {{\Psi}  {\Psi} }} \right.
\kern-\nulldelimiterspace} {{\Psi} } \right\rangle} }  , \nonumber
\\
p_{zz}&=&\frac{{\left( {a^{2} + c^{2}} \right)e^{ - \rho \sigma
_{1} z} - 2b^{2}e^{ - \rho \sigma _{0} z}}}{{\left\langle {{\Psi}
} \mathrel{\left| {\vphantom {{\Psi} {\Psi} }} \right.
\kern-\nulldelimiterspace} {{\Psi} }
\right\rangle} } ,\nonumber\\
p_{xy}&=& 0 ,\nonumber
\\
p_{xz}&=&\frac{{ \frac{{3}}{{\sqrt {2}} }e^{ - \frac{{1}}{{2}}\rho
\left( {\sigma _{0} + \sigma _{1}} \right)z}b\left( {a + c}
\right)\sin\left( {\frac{{2\pi \rho} }{{k}}\Re d_{1} z}
\right)}}{{\left\langle {{\Psi} } \mathrel{\left| {\vphantom
{{\Psi}  {\Psi }}} \right. \kern-\nulldelimiterspace} {{\Psi} }
\right\rangle} }  ,\nonumber
\\
p_{yz}&=&\frac{{\frac{{3}}{{\sqrt {2}} }e^{ - \frac{{1}}{{2}}\rho
\left( {\sigma _{0} + \sigma _{1}} \right)z}b\left( {a - c}
\right)\cos\left( {\frac{{2\pi \rho} }{{k}}\Re d_{1} z}
\right)}}{{\left\langle {{\Psi} } \mathrel{\left| {\vphantom
{{\Psi}  {\Psi }}} \right. \kern-\nulldelimiterspace} {{\Psi} }
\right\rangle} } , \label{rot2}
\end{eqnarray}
\\
\noindent where $\left\langle {{\Psi} } \mathrel{\left| {\vphantom
{{\Psi} {\Psi} }} \right. \kern-\nulldelimiterspace} {{\Psi} }
\right\rangle = \left( {a^{2} + c^{2}} \right)e^{ - \rho \sigma
_{1} z} + b^{2}e^{ - \rho \sigma _{0} z}$,
$\sigma_{0}=\frac{{4\pi} }{{k}}\Im f_0$, $\sigma _{1} =
\frac{{4\pi} }{{k}}\Im f_1$, $f_0=d$, $f_1=d+d_1$.

According to (\ref{rot1}), (\ref{rot2}) spin rotation occurs when
the angle between the polarization vector $\vec{p}$ and momentum
$\vec{k}$ of the particle differs from $\frac{{\pi} }{{2}}$.

For example, when $\Re d_1 >0$, the angle between the polarization
vector and momentum is acute (see Fig.\ref{angle1}) and the spin
rotates left-hand around the momentum direction, whereas an obtuse
angle between the polarization vector and the momentum gives rise
to right-hand spin rotation (see Fig.\ref{angle2}).
\begin{center}
\begin{figure}[!h]
\epsfxsize = 5 cm \centerline{\epsfbox{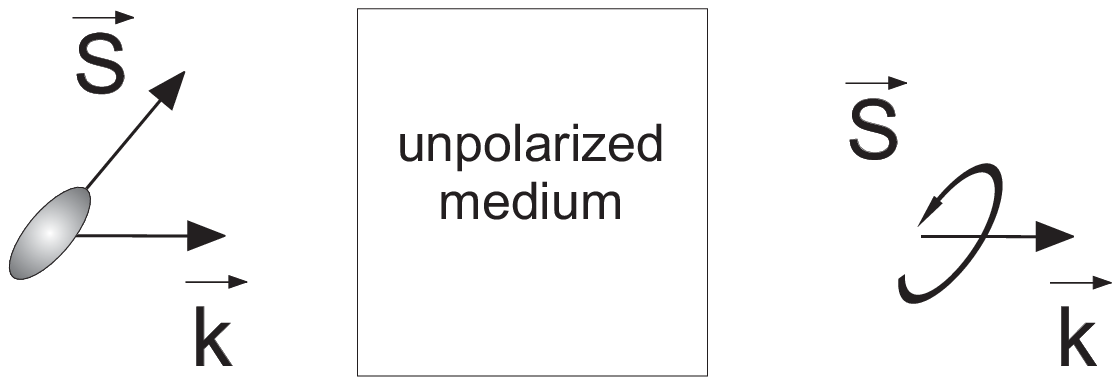}}
\caption{}
 \label{angle1}
\end{figure}
\begin{figure}[!h]
\epsfxsize = 5 cm \centerline{\epsfbox{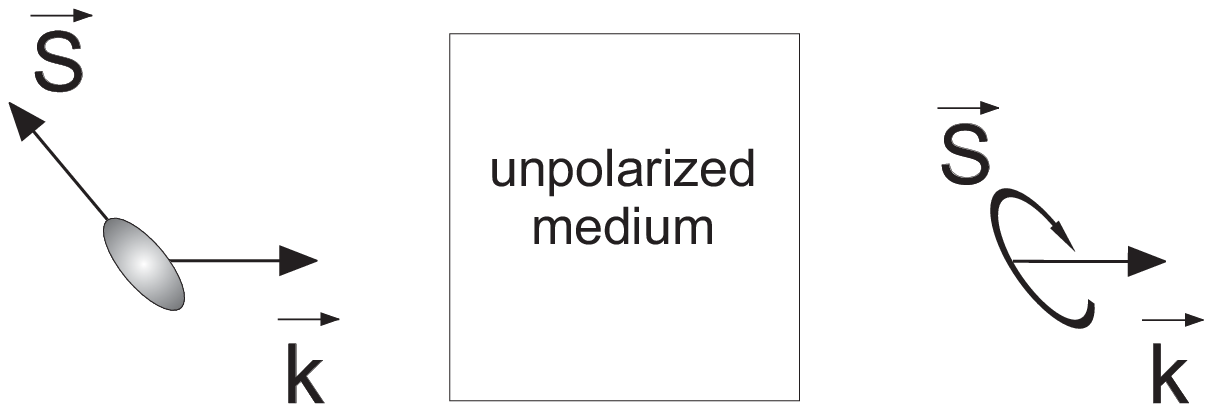} }
\caption{}
 \label{angle2}
\end{figure}
\end{center}
%------------------------------------------------------------------------------insert figure angle1,2
%
%
When the polarization vector and momentum are perpendicular
 (a transversely polarized particle), then  the components of
 the vector polarization at $z = 0$ are: $p_x= 0$, $p_y\ne0$, and $p_z=0$.
In this case $a=c$ and dependance of the vector polarization on
$z$ can be expressed as:
\begin{eqnarray}
p_x&=&0,\nonumber\\
p_y&=&\frac{{\sqrt {2} e^{ - \frac{{1}}{{2}}\rho \left( {\sigma
_{0} + \sigma _{1}} \right)z}2ba\cos\left( {\frac{{2\pi \rho
}}{{k}}\Re d_{1} z} \right)}}{{\left\langle {{\Psi} }
\mathrel{\left| {\vphantom {{\Psi}  {\Psi} }} \right.
\kern-\nulldelimiterspace} {{\Psi} } \right\rangle} },\nonumber\\
p_z&=&0,\nonumber\\
p_{xx}&=&\frac{{ - 4a^{2}e^{ - \rho \sigma _{1} z} + b^{2}e^{ -
\rho \sigma _{0} z}}}{{\left\langle {{\Psi} } \mathrel{\left|
{\vphantom {{\Psi}  {\Psi} }} \right.
\kern-\nulldelimiterspace} {{\Psi} } \right\rangle} }, \\
p_{yy}&=&\frac{{2a^{2}e^{ - \rho \sigma _{1} z} + b^{2}e^{ - \rho
\sigma _{0} z}}}{{\left\langle {{\Psi} } \mathrel{\left|
{\vphantom {{\Psi}  {\Psi} }} \right.
\kern-\nulldelimiterspace} {{\Psi} } \right\rangle} },\nonumber\\
p_{zz}&=&\frac{{2a^{2}e^{ - \rho \sigma_{1} z} - 2b^{2}e^{ - \rho
\sigma _{0} z}}}{{\left\langle {{\Psi} } \mathrel{\left|
{\vphantom {{\Psi}  {\Psi} }} \right. \kern-\nulldelimiterspace}
{{\Psi} } \right\rangle} }, \nonumber
\\
p_{xz}&=&\frac{{  \frac{{3}}{{\sqrt {2}} }e^{ -
\frac{{1}}{{2}}\rho \left( {\sigma _{0} + \sigma _{1}}
\right)z}2ab\sin\left( {\frac{{2\pi \rho} }{{k}}\Re d_{1} z}
\right)}}{{\left\langle {{\Psi} } \mathrel{\left| {\vphantom
{{\Psi}  {\Psi }}} \right. \kern-\nulldelimiterspace} {{\Psi} }
\right\rangle} },\nonumber\\
p_{yz}&=&0. \nonumber \label{11}
\end{eqnarray}
%
%--- figure oscillate ---------------------
%
According to (\ref{11}) the vector and tensor polarization
oscillate when a transversely polarized deuteron passes through
matter (see Fig.\ref{oscillate})
 i.e. the vector
polarization converts to the tensor one and vice versa (similar
the linear to circular polarization conversion for a photon in an
optically anisotropic medium - the well known optical
birefringence). However, unlike the optical birefringence, the
birefringence effect for particles appear in isotropic matter
 (and even the spin of matter nuclei is either zero or unpolarized !). Anisotropy is
 provided by the particle itself
 (a particle with the spin $S \ge 1$ and mass $M \ne 0$ has the intrinsic anisotropy).
\begin{center}
\begin{figure}[!h]
\epsfxsize = 5 cm \centerline{\epsfbox{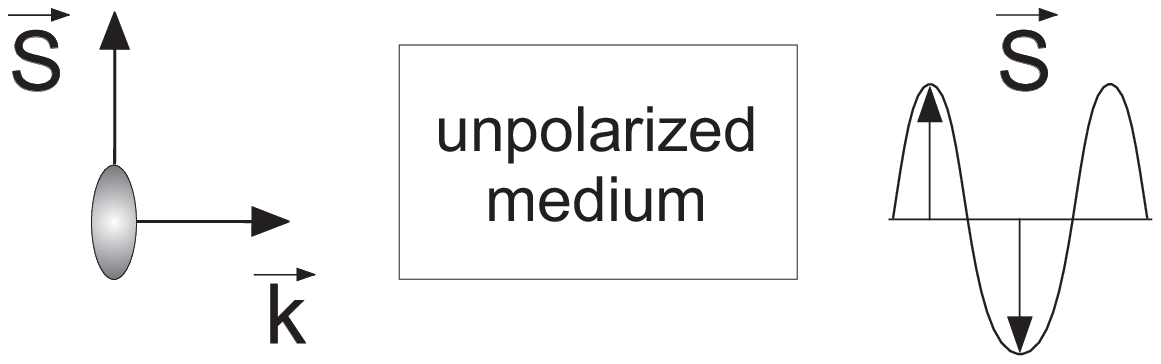}}
 \caption{}
 \label{oscillate}
\end{figure}
\end{center}
%---------------------

% ---------------------------  end from file Cosy experiment for PR 2005-03-31.tex ------------------
\bigskip

%-------- to be changed !!!!!!!!!!!!!!!!!!!!!!!!!!!!!!!!!!!!!!!!!!!!!!!!!!!-------------

According to the above (see (\ref{U1})) a particle possesses some
effective potential energy $V$ in matter. If the particle spin $S
\ge 1$, then this energy depends on the spin orientation
\cite{1,2,4}
\begin{equation}
\hat{V}=-\frac{2\pi \hbar^{2}}{M \gamma}N\hat{f(0)},
 \label{1.1}
\end{equation}
 where $M$ is the particle mass, $\hat{f(0)}$ is the
spin dependent zero-angle elastic coherent scattering amplitude of
the particle, $N$ is the density of the scatterers in the medium
(the number of scatterers in $1cm^{3}$), $\gamma$ is the Lorentz
factor.
 Substituting $\hat{f(0)}$ for a particle with the spin
 $S=1$  in (\ref{1.1}) in the explicit form one can obtain \cite{1,2,4}
\begin{equation}
\hat{V}=-\frac{2\pi
\hbar^{2}}{M\gamma}N\left(d+d_{1}\left(\vec{S}\vec{n}\right)^{2}\right),
\label{1.2}
\end{equation}
 where  $\vec{n}$ is the unit vector along the particle momentum
direction.

Let the quantization axis z is directed along $\vec{n}$ and $m$
denotes the magnetic quantum number. Then, for a particle in a
state that is an eigenstate of the operator $S_{z}$ of spin
projection onto the z-axis, the effective potential energy can be
written as:
\begin{equation}
\hat{V}=-\frac{2\pi \hbar^{2}}{M\gamma}N\left(d+d_{1}m^{2}\right).
\label{1.3}
\end{equation}

 According to (\ref{1.3}) splitting of the deuteron energy levels
in matter is similar to splitting of atom energy levels in an
electric field aroused from the quadratic Stark effect.
 Therefore,
the above effect could be considered as caused by splitting of the
spin levels of the particle in the pseudoelectric nuclear field of
matter.

Let an electric field  acts on a deuteron (nucleus). The energy
$\hat{V}_{E}$ of the deuteron beam in the external electric field
$\vec{E}$ due to the tensor electric polarizability can be written
in the form
\begin{equation}
\hat{V}_{E}=-\frac{1}{2}\hat{\alpha}_{ik}E_{i}E_{k},
 \label{1.4}
\end{equation}
where $\hat{\alpha}_{ik}$ is the deuteron tensor electric
polarizability, $E_{i}$ are the components of the electric field.
This expression can be rewritten as follows:
\begin{equation}
\hat{V}_{E}=\alpha_{S}E^{2}-\alpha_{T}E^{2}\left(\vec{S}\vec{n}_{E}\right)^{2},
%\eqno(1.5)
 \label{1.5}
\end{equation}
where $\alpha_{S}$ is the deuteron scalar electric polarizability,
$\alpha_{T}$ is the deuteron tensor electric polarizability,
$\vec{n}_{E}$ is the unit vector along $\vec{E}$.

 {Comparison of (\ref{1.5}) with (\ref{1.2}) provides to conclude that in an
electric field we can observe the effect of spin rotation and
oscillations about the $\vec{E}$ direction for a particle with $S
\ge 1$ , too \cite{7}.}

\section{THE EQUATIONS FOR THE POLARIZATION VECTOR AND
QUADRUPOLARIZATION TENSOR OF THE DEUTERON BEAM IN A STORAGE RING}

 Considering evolution of the spin of a particle in a
storage ring one should take into account several interactions:

1. interactions of the magnetic and electric dipole moments with
an electromagnetic field;

2. interaction (\ref{1.5}) of the particle with the  electric
field due to the tensor electric polarizability

3. interaction (\ref{1.2}) of the particle with the pseudoelectric
nuclear field of matter.

Therefore, the equation for the particle spin wavefunction is:
\begin{equation}
i\hbar\frac{\partial\Psi(t)}{\partial
t}=\left(\hat{H}_{0}+\hat{V}_{d}+\hat{V}+\hat{V}_{E}\right)\Psi(t)
%\eqno(1.6)
\label{1.6}
\end{equation}
where $\Psi(t)$ is the particle spin wavefunction,
{$\hat{H}_{0}$ is the Hamiltonian describing the spin behavior
caused by interaction of the magnetic moment with the
electromagnetic field (equation (\ref{1.6}) with the only
$\hat{H}_{0}$  summand converts to the Bargman-Myshel-Telegdy
equation),}
 $\hat{V}_{d}$ describes interaction of the deuteron (nuclear)
EDM with the electric field.

%%%%%%%%%%%%%%%%%%%%%

Let us describe motion of a deuteron in a storage ring in external
magnetic and electric fields. Particle spin precession induced by
interaction of the magnetic moment of a particle with an external
electromagnetic field  can be described by the
Bargman-Myshel-Telegdy equation \cite{6,8}
\begin{equation}
\frac{d\vec{p}}{dt}=[\vec{p}\times\vec{\Omega}_{0}],
%\eqno(2.1)
\label{2.1}
\end{equation}
where $t$ is time in the laboratory system,
\begin{equation}
\vec{\Omega}_{0}=\frac{e}{mc}\left[\left(a+\frac{1}{\gamma}\right)\vec{B}
-a\frac{\gamma}{\gamma+1}\left(\vec{\beta}\cdot\vec{B}\right)\vec{\beta}-
\left(\frac{g}{2}-\frac{\gamma}{\gamma+1}\right)\vec{\beta}\times\vec{E}\right],
%\eqno(2.2)
\label{2.2}
\end{equation}
$m$ is the mass of the particle, $e$ is its charge, $\vec{p}$ is
the spin polarization vector, $\gamma$ is the Lorentz-factor,
 $\vec{\beta}=\vec{v}/c$, $\vec{v}$ is the particle velocity, $a=(g-2)/2$, $g$ is the gyromagnetic ratio, $\vec{E}$ and
 $\vec{H}$ are the electric and magnetic fields in the point of
 particle location.

If a particle possesses an intrinsic dipole moment then the
additional term that describes the spin rotation induced by the
EDM should be added to (\ref{2.1}) \cite{6}
\begin{equation}
\frac{d\vec{p}_{edm}}{dt}=\frac{d}{\hbar}
\left[\vec{p}\times\left(\vec{\beta}\times\vec{B}+\vec{E}\right)\right],
%\eqno(2.3)
\label{2.3}
\end{equation}
where $d$ is the electric dipole moment of a particle.

As a result, evolution of the deuteron spin due to the magnetic
and electric momenta can be described by the following equation:
\begin{eqnarray}
\frac{d\vec{p}}{dt}=
\frac{e}{mc}\left[\vec{p}\times\left\{\left(a+\frac{1}{\gamma}\right)\vec{B}
-a\frac{\gamma}{\gamma+1}\left(\vec{\beta}\cdot\vec{B}\right)\vec{\beta}-
\left(\frac{g}{2}-\frac{\gamma}{\gamma+1}\right)\vec{\beta}\times\vec{E}\right\}\right]
+d\left[\vec{p}\times\left(c\vec{\beta}\times\vec{B}+\vec{E}\right)\right].
%\eqno(2.4)
\label{2.4}
\end{eqnarray}

According to the section 2, the equation (\ref{2.4}) does not
describe the particle spin evolution in a storage ring completely.
 The expression (\ref{2.4}) should be
supplemented with the additions given by interactions
$\hat{V}_{E}$ and $\hat{V}$ (see (\ref{1.2}-\ref{1.6})) .

This additional contribution could be found by the aids of the
particle spin wavefunction $\Psi(t)$ (see \ref{1.6})).

 The equations describing the time evolution of the spin
and tensor polarization {caused by the phenomena of birefringence}
can be written as:
\begin{eqnarray}
\frac{d\vec{p}}{dt}& = &
\frac{d}{dt}\frac{\langle\Psi(t)|\vec{S}|\Psi(t)\rangle}{\langle\Psi(t)|\Psi(t)|},
\nonumber \\
\frac{dp_{ik}}{dt} & = &
\frac{d}{dt}\frac{\langle\Psi(t)|Q_{ik}|\Psi(t)\rangle}{\langle\Psi(t)|\Psi(t)|},
%\eqno(2.5)
\label{2.5}
\end{eqnarray}
where $\Psi(t)$ is the deuteron wave function,
$\hat{Q}_{ik}=\frac{3}{2}\left(S_{i}S_{k}+S_{k}S_{i}-\frac{4}{3}\delta_{ik}\hat{\texttt{I}}\right)$
is the tensor of rank two (the tensor polarization).

 The equations (\ref{2.5}) contain initial phases that
determine the deuteron wave function.
 Therefore, a partly
polarized beam can not be described by such equations.
 So the
 spin density matrix formalism should be used to derive
equations describing the evolution of the deuteron spin
\cite{nastya}.

The density matrix of the system "deuteron+target" is
\begin{eqnarray}
\rho=\rho_{d}\otimes\rho_{t},
%\eqno(2.6)
 \label{2.6}
\end{eqnarray}
where $\rho_{d}$ is the density matrix of the deuteron beam
\begin{eqnarray}
\rho_{d}=I(\vec{k})\left(\frac{1}{3}\hat{\texttt{I}}
+\frac{1}{2}\vec{p}(\vec{k})\vec{S}+\frac{1}{9}p_{ik}(\vec{k})\hat{Q}_{ik}\right),
%\eqno(2.7)
\label{2.7}
\end{eqnarray}
$I(\vec{k})$ is the intensity of the beam, $\vec{p}$ is the
polarization vector, $p_{ik}$ is the tensor polarization of the
deuteron beam, $\rho_{t}$ is the density matrix of the target.
{For an unpolarized target $\rho_{t}=\hat{\texttt{I}}$, where
$\hat{\texttt{I}}$ is the unit matrix in the spin space of target
particle.}

The equation for the deuteron beam density matrix can be written
as:
\begin{eqnarray}
\frac{d\rho_{d}}{dt}=-\frac{i}{\hbar}\left[\hat{H},\rho_{d}\right]+\left(\frac{\partial\rho_{d}}{\partial
t}\right)_{col},
%\eqno(2.9)
\label{2.9}
\end{eqnarray}
where $\hat{H}=\hat{H}_{0}+\hat{V}_{d}+\hat{V}_{E}$,
\begin{eqnarray}
\hat{V}_{d} & = &
-d\left(\vec{\beta}\times\vec{B}+\vec{E}\right)\vec{S},
%\eqno(2.10)
\label{2.10}\\
\hat{V}_{E} & = &
\alpha_{S}\left(\vec{\beta}\times\vec{B}+\vec{E}\right)^{2}
-\alpha_{T}\left(\vec{\beta}\times\vec{B}+\vec{E}\right)^{2}\left(\vec{S}\vec{n}_{E}\right)^{2},
\nonumber \\
\vec{n}_{E} & = &
\frac{\vec{E}+\vec{\beta}\times\vec{B}}{|\vec{E}+\vec{\beta}\times\vec{B}|}.
\nonumber
\end{eqnarray}
 The collision term
$\left(\frac{\partial\rho_{d}}{\partial t}\right)_{col}$ can be
found by the method described in \cite{9}:
\begin{eqnarray}
\left(\frac{\partial\rho_{d}}{\partial
t}\right)_{col}
=vN\emph{Sp}_{t}\left[\frac{2\pi
i}{k}\left[F(\theta=0)\rho-\rho F^{+}(\theta=0)\right] +\int
d\Omega F(\vec{k}^{'})\rho(\vec{k}^{'})F^{+}(\vec{k}^{'})\right],
%\eqno(2.11)
\label{2.11}
\end{eqnarray}
where $\vec{k}^{'}=\vec{k}+\vec{q}$, {$\vec{q}$ is the momentum
carried over a nucleus of the matter from the incident particle,}
 $v$ is the speed of the incident particles, $N$ is the
atom density in the matter,
 $F$ is the scattering amplitude
depending on the spin operators of the deuteron and the matter
nucleus (atom), $F^+$ is the Hermitian conjugate of the operator
$F$.
{The first term in (\ref{2.11}) describes coherent scattering of a
particle by matter nuclei, while the second term is for multiple
scattering.}

Let us consider the first term in (\ref{2.11}):
\begin{eqnarray}
\left(\frac{\partial\rho_{d}}{\partial
t}\right)_{col}^{(1)}=vN\frac{2\pi i}{k}
 \left[
 \hat{f}(0)\rho_d-\rho_d \hat{f}(0)^{+}
\right] .
%\eqno(2.12)
\label{2.12}
\end{eqnarray}
The  amplitude $\hat{f}(0)$ of deuteron scattering in an
unpolarized target at the zero angle is
\begin{eqnarray}
\hat{f}(0)=\emph{Sp}_{t}F(0)\rho_{t}.
%\eqno(2.13)
\label{2.13}
\end{eqnarray}
This amplitude can de rewritten according to (\ref{1.2}) as
\begin{eqnarray}
\hat{f}(0)=d+d_{1}(\vec{S}\vec{n})^{2},
%\eqno(2.14)
\label{2.14}
\end{eqnarray}
where $\vec{n}=\vec{k}/k$, $\vec{k}$ is the deuteron momentum.

As a result one can obtain:
\begin{eqnarray}
\left(\frac{\partial\rho_{d}}{\partial t}\right)_{col}^{(1)} =
 -\frac
i\hbar\left(\hat{V} {\rho_d}-{\rho_d} \hat{V}^{+}\right).
%\eqno(2.15)
\label{2.15}
\end{eqnarray}

Finally, the expression (\ref{2.9}) reads
\begin{eqnarray}
\frac{d\rho_{d}}{dt}=-\frac{i}{\hbar}\left[\hat{H},\rho_{d}\right]
 -\frac
i\hbar\left(\hat{V} {\rho_d}-{\rho_d} \hat{V}^{+}\right)+
 vN
\emph{Sp}_{t} \int d\Omega
F(\vec{k}^{'})\rho(\vec{k}^{'})F^{+}(\vec{k}^{'}).
 %\eqno(2.9)
 \label{2.9_new}
\end{eqnarray}
The last term in the above formula, which is proportional to
$\emph{Sp}_{t}$, describes the multiple scattering process and
spin depolarization aroused from it. Henceforward we consider such
time of experiment (such effective length for a particle in a
matter) that provides to neglect this term.

The intensity of the beam is
\begin{eqnarray}
I=\emph{Sp}_{d}\rho_{d}.
% \eqno(2.16)
\label{2.16}
\end{eqnarray}
Consequently
\begin{eqnarray}
\frac{dI}{dt}=vN\frac{2\pi
i}{k}\emph{Sp}_{d}\left[f(0)\rho_{d}-\rho_{d}f^{+}(0)\right].
%\eqno(2.17)
\label{2.17}
\end{eqnarray}
Substituting (\ref{2.7}) and (\ref{2.14}) into (\ref{2.17}) we can
get
\begin{eqnarray}
\frac{dI}{dt}=\frac{\chi}{3}\left[2+p_{ik}n_{i}n_{k}\right]I(t)+\alpha
I(t),
%\eqno(2.18)
\label{2.18}
\end{eqnarray}
where $\chi=-\frac{4\pi
vN}{k}\texttt{Im}d_{1}=--vN(\sigma_1-\sigma_0)$,
$\alpha=-\frac{4\pi vN}{k}\texttt{Im}d=-v N \sigma_0$, $\sigma_1$
and $\sigma_0$ are the total cross-sections of deuteron scattering
by a nonpolarized nucleus for the magnetic quantum numbers $m=1$
and $m=0$, respectively.

The vector polarization  of the deuteron beam $\vec{p}$ is
determined as
\begin{eqnarray}
\vec{p}=\frac{\emph{Sp}_{d}\rho_{d}\vec{S}}{\emph{Sp}_{d}\rho_{d}}=\frac{\emph{Sp}_{d}\rho_{d}\vec{S}}{I}.
%\eqno(2.19)
\label{2.19}
\end{eqnarray}

From (\ref{2.19}) one can get the differential equation for the
beam polarization
\begin{eqnarray}
\frac{d\vec{p}}{dt}=\frac{\emph{Sp}_{d}(d\rho_{d}/dt)\vec{S}}{I(t)}-
\vec{p}\frac{\emph{Sp}_{d} (d\rho_{d}/dt)}{I(t)}.
%\eqno(2.20)
\label{2.20}
\end{eqnarray}

The expression for the tensor polarization is
\begin{eqnarray}
p_{ik}=\frac{\emph{Sp}_{d}\rho_{d}Q_{ik}}{\emph{Sp}_{d}\rho_{d}}=\frac{\emph{Sp}_{d}\rho_{d}Q_{ik}}{I},
%\eqno(2.21)
\label{2.21}
\end{eqnarray}
where
$\hat{Q}_{ik}=\frac{3}{2}\left(S_{i}S_{k}+S_{k}S_{i}-\frac{4}{3}\delta_{ik}\hat{\texttt{I}}\right)$.

The change of the tensor polarization can be written as
\begin{eqnarray}
\frac{dp_{ik}}{dt}=\frac{\emph{Sp}_{d}(d\rho_{d}/dt)Q_{ik}}{I(t)}-
p_{ik}\frac{\emph{Sp}_{d} (d\rho_{d}/dt)}{I(t)}.
%\eqno(2.22)
\label{2.22}
\end{eqnarray}

Using (\ref{2.7}) and (\ref{2.1}), (\ref{2.20}) and (\ref{2.22})
we can get the equation system  for the time evolution of the
deuteron polarization vector and quadrupolarization tensor
($\vec{n}=\vec{k}/k$,
$\vec{n}_{E}=\frac{\vec{E}+\vec{\beta}\times\vec{B}}{|\vec{E}+\vec{\beta}\times\vec{B}|}$,
$p_{xx}+p_{yy}+p_{zz}=0$) \cite{kostya}
\begin{eqnarray}
\left\{
\begin{array}{l}
\frac{d\vec{p}}{dt}=
\frac{e}{mc}\left[\vec{p}\times\left\{\left(a+\frac{1}{\gamma}\right)\vec{B}
-a\frac{\gamma}{\gamma+1}\left(\vec{\beta}\cdot\vec{B}\right)\vec{\beta}-
\left(\frac{g}{2}-\frac{\gamma}{\gamma+1}\right)\vec{\beta}\times\vec{E}\right\}\right]+\\
 +
\frac{d}{\hbar}\left[\vec{p}\times\left({\vec{E}}+\vec{\beta}\times\vec{B}\right)\right]
+\frac{\chi}{2}(\vec{n}(\vec{n}\cdot\vec{p})+\vec{p}) + \\
 +  \frac{\eta}{3}[\vec{n}\times\vec{n}^{'}]
-\frac{2\chi}{3}\vec{p}-\frac{\chi}{3}(\vec{n}\cdot\vec{n}^{'})\vec{p}
-\frac{2}{3}\frac{\alpha_{T}E^{2}_{eff}}{\hbar}[\vec{n}_{E}\times\vec{n}_{E}^{'}],\\
{} \\
\frac{dp_{ik}}{dt}  =  -\left(\varepsilon_{jkr}p_{ij}\Omega_{r}+\varepsilon_{jir}p_{kj}\Omega_{r}\right) + \\
 +
\chi\left\{-\frac{1}{3}+n_{i}n_{k}+\frac{1}{3}p_{ik}-\frac{1}{2}(n_{i}^{'}n_{k}+n_{i}n_{k}^{'})
+\frac{1}{3}(\vec{n}\cdot\vec{n}^{'})\delta_{ik}\right\} + \\
 +
\frac{3\eta}{4}\left([\vec{n}\times\vec{p}]_{i}n_{k}+n_{i}[\vec{n}\times\vec{p}]_{k}\right)
-\frac{\chi}{3}(\vec{n}\cdot\vec{n}^{'})p_{ik} - \\
 -
\frac{3}{2}\frac{\alpha_{T}E^{2}_{eff}}{\hbar}\left([\vec{n}_{E}\times\vec{p}]_{i}n_{E,\,k}
+n_{E,\,i}[\vec{n}_{E}\times\vec{p}]_{k}\right),
\\
\end{array}
\right.
%\eqno(2.23)
\label{2.23}
\end{eqnarray}
where $\vec{E}_{eff}=(\vec{E}+\vec{\beta} \times \vec{B})$,
$\eta=-\frac{4 \pi N}{k} \texttt{Re}d_{1}$,
$n_{i}^{'}=p_{ik}n_{k}$, $n_{E,\,i}^{'}=p_{ik}n_{E,\,k}$\,,
$\Omega_{r}(d)$ are the components of the vector $\vec{\Omega}(d)$
($r=1,2,3$ corresponds $x,y,z$):
\begin{eqnarray}
\vec{\Omega}(d) & = &
\frac{e}{mc}\left\{\left(a+\frac{1}{\gamma}\right)\vec{B}
-a\frac{\gamma}{\gamma+1}\left(\vec{\beta}\cdot\vec{B}\right)\vec{\beta}-
\left(\frac{g}{2}-\frac{\gamma}{\gamma+1}\right)\vec{\beta}\times\vec{E}\right\} + \nonumber \\
& + &
\frac{d}{\hbar}\left({\vec{E}}+\vec{\beta}\times\vec{B}\right).
%\eqno(2.24)
\label{2.24}
\end{eqnarray}

Then we consider the spin rotation about the particle momentum.
 According to \cite{6,project} the spin precession caused by the magnetic
moment ($(g-2)$ precession) can be minimized and even zeroized by
applying a radial electric field.

{The angles of spin rotation caused by both the EDM and
birefringence effect are small for the considered experiment
duration.}
 Therefore, the perturbation theory can be used for (\ref{2.23})
 solution.
\begin{eqnarray}
\vec{p}(t)=\vec{p}^{\,0}+\frac{e}{mc}\left[\vec{p}^{\,0}\times\left\{a\vec{B}
+\left(\frac{1}{\gamma^{2}-1}-a\right)\vec{\beta}\times\vec{E}\right\}\right]t
+
\nonumber \\
+\frac{d}{\hbar}\left[\vec{p}^{\,0}\times\left({\vec{E}}+\vec{\beta}\times\vec{B}\right)\right]t
+ \{
\frac{\chi}{2}(\vec{n}(\vec{n}\cdot\vec{p}^{\,0})+\vec{p}^{\,0})t
+\frac{\eta}{3}[\vec{n}\times\vec{n}^{'}_{0}]t - \nonumber \\
-\frac{2\chi}{3}\vec{p}^{0}t
-\frac{\chi}{3}(\vec{n}\cdot\vec{n}^{'}_{0})\vec{p}^{\,0}t \}
-\frac{2}{3}\frac{\alpha_{T}E^{2}_{eff}}{\hbar}[\vec{n}_{E}\times\vec{n}_{E\,0}^{'}]t,
%\eqno(2.25)
\label{2.25}
\end{eqnarray}
\begin{eqnarray}
\lefteqn
p_{ik}(t)  =
p_{ik}^{0}-\frac{e}{mc}(\varepsilon_{jkr}p_{ij}+\varepsilon_{jir}p_{kj})\left\{a\vec{B}
+\left(\frac{1}{\gamma^{2}-1}-a\right)\vec{\beta}\times\vec{E}\right\}_{r}t - \nonumber\\
-
\frac{d}{\hbar}(\varepsilon_{jkr}p_{ij}+\varepsilon_{jir}p_{kj})\left({\vec{E}}
+\vec{\beta}\times\vec{B}\right)_{r}t + \nonumber \\
+ \{
 \chi\left[-\frac{1}{3}+n_{i}n_{k}+\frac{1}{3}p_{ik}^{0}
-\frac{1}{2}(n_{i0}^{'}n_{k}+n_{i}n_{k0}^{'})
+\frac{1}{3}(\vec{n}\cdot\vec{n}^{'}_{0})\delta_{ik}\right]t + \nonumber \\
 +  \frac{3\eta}{4}\left([\vec{n}\times\vec{p}^{\,0}]_{i}n_{k}
+n_{i}[\vec{n}\times\vec{p}^{\,0}]_{k}\right)t
-\frac{\chi}{3}(\vec{n}\cdot\vec{n}^{'}_{0})p_{ik}^{\,0}t \}-
\nonumber \\
-\frac{3}{2}\frac{\alpha_{T}E^{2}_{eff}}{\hbar}\left([\vec{n}_{E}\times\vec{p}^{\,0}]_{i}n_{E,\,k}
+n_{E,\,i}[\vec{n}_{E}\times\vec{p}^{\,0}]_{k}\right)t,
%\eqno(2.26)
\label{2.26}
\end{eqnarray}
where $\vec{p}^{\,0}$ is the beam polarization at $t_{0}=0$,
$n_{i0}^{'}=p_{ik}^{\,0}n_{k}$,
$n_{E\,0,\,i}^{'}=p_{ik}^{\,0}n_{E,\,k}$, $p_{ik}^{\,0}$ are the
components of the tensor polarization at the initial moment of
time.

In real situation, even when $(g-2)$ precession is suppressed,
nevertheless, the rotation angle can appear large enough (during
the experiment the spin can rotate several turns \cite{project}).
Absorption can also appear significant. In this case one should
analyze the system (\ref{2.23}) instead of perturbation theory
results (\ref{2.25}).

Thus, according to (\ref{2.25}), (\ref{2.26}) the  spin behavior
of a deuteron rotating in a storage ring is caused by several
contributions:

\noindent 1. spin rotation which is described by
Bargman-Myshel-Telegdy equation;

\noindent 2. rotation due to the deuteron EDM;

\noindent 3. rotation and dichroism due to the the phenomena of
birefringence in matter;

\noindent and

\noindent 4. spin rotation due to the phenomena of birefringence
in an electric field.

Let us consider some particular cases.

%\textbf{2.1}
\textbf{Case I.}
% Let us consider the beam spin state behavior due to
%the EDM and the phenomenon of birefringence in a medium. If at
%$t_{0}=0$ the beam spin state is described by the direction of
Suppose the vector polarization is parallel to the the $z$ axis,
i.e. $p_{x}^{\,0}=p_{y}^{\,0}=0$, $p_{z}^{\,0}\neq0$,
$p_{ik}^{\,0}=0$, if  $i\neq k$,  $p_{xx}\neq 0$,
$p_{yy}^{\,0}\neq 0$, $p_{zz}^{\,0}=0$.

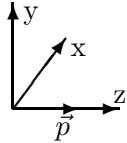
\begin{figure}[h]
\begin{center}
\begin{picture}(40,40)\thicklines
\put(0,0){\vector(1,0){40}} \put(0,0){\vector(0,1){40}}
\put(0,0){\vector(3,4){20}} \put(38,2){z} \put(4,34){y}
\put(22,20){x} \put(0,0){\vector(1,0){25}} \put(16,-10){$\vec{p}$}
\end{picture}
\end{center}
\caption{The original orientation of the polarization vector}
\label{Fig.1}
\end{figure}

Suppose $(g-2)$ spin precession caused by the magnetic moment is
zero, then one can obtain
\begin{eqnarray}
p_{x}(t)=0 \nonumber\\
p_{y}(t)=-\frac{d}{\hbar}\,p_{z}^{\,0}\left({E}+\beta
B\right)t , \nonumber\\
p_{z}(t)=p_{z}^{\,0}+\frac{1}{3}\chi p_{z}^{\,0}t , \nonumber\\
p_{xx}(t)=p_{xx}^{\,0}+\frac{\chi}{3}\left(-1+p_{xx}^{\,0}\right)t , \nonumber\\
p_{yy}(t)=p_{yy}^{\,0}+\frac{\chi}{3}\left(-1+p_{yy}^{\,0}\right)t , \\
p_{zz}(t)=\frac{2\chi}{3}t , \nonumber\\
p_{xy}(t)=p_{xz}(t)=0 , \nonumber\\
p_{yz}(t)=\frac d \hbar \,p_{yy}\left(E +\beta B\right)t, \nonumber\\
p_{xy}=\frac{3}{2}\frac{\alpha_{T}E^{2}_{eff}}{\hbar}p_{z}^{\,0}
\nonumber
%\eqno(2.27)
\label{2.27}
\end{eqnarray}
The solution (\ref{2.27}) shows that
 even in this practically ideal case (when the polarization vector is exactly parallel to $\vec{n}$)
 change of polarization due to the birefringence effect leads to the
 appearance of additional components of $\vec{p}$ and $p_{ik}$
 along with the components aroused
 from the deuteron EDM.

 %\textbf{2.2}
 Case II.
 Let us consider now the more real case.

\begin{figure}[h]
\begin{center}
\begin{picture}(40,40)\thicklines
\put(0,0){\vector(1,0){40}} \put(0,0){\vector(0,1){40}}
\put(0,0){\vector(3,4){20}} \put(38,2){z} \put(4,34){y}
\put(22,20){x} \put(0,0){\vector(4,1){28}} \put(24,8){$\vec{p}$}
\end{picture}
\end{center}
\caption{The original orientation of the polarization vector}
\label{Fig.2}
\end{figure}
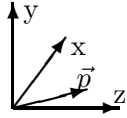

Suppose the angle between the initial polarization vector and the
$z$ axis is acute (Fig.\ref{Fig.2})
Let us express the effect caused by the gas inside the storage
ring.
 Then the solution of (\ref{2.25},\ref{2.26}) can be written
as follows:
\begin{eqnarray}
p_{x}(t)=p_{x}^{\,0}-\frac{\chi}{6}\,p_{x}^{\,0}\,t-\frac{\chi}{3}\,p_{zz}^{\,0}\,p_{x}^{\,0}\,t
-\frac{\eta}{3}\,p_{yz}^{\,0}t \nonumber \\
p_{y}(t)=p_{y}^{\,0}-\frac{d}{\hbar}\left({E}+\beta
B\right)p_{z}^{\,0}\,t-\frac{\chi}{6}\,p_{y}^{\,0}\,t
-\frac{\chi}{3}\,p_{zz}^{\,0}\,p_{y}^{\,0}\,t+\frac{\eta}{3}\,p_{xz}^{\,0}t \nonumber \\
p_{z}(t)=p_{z}^{\,0}+\frac{d}{\hbar}\left({E}+\beta
B\right)p_{y}^{\,0}\,t+\frac{\chi}{3}\,p_{z}^{\,0}\,t
-\frac{\chi}{3}\,p_{zz}^{\,0}\,p_{z}^{\,0}\,t\nonumber \\
p_{xx}(t)=p_{xx}^{\,0}-\frac{\chi}{3}\left(1+p_{yy}^{\,0}\right)\,t
-\frac{\chi}{3}\,p_{zz}^{\,0}\,p_{xx}^{\,0}\,t\nonumber \\
p_{yy}(t)=p_{yy}^{\,0}-2\frac d\hbar\,p_{yz}\left(E+\beta
B\right)-\frac{\chi}{3}\left(1+p_{xx}^{\,0}\right)\,t
-\frac{\chi}{3}\,p_{zz}^{\,0}\,p_{yy}^{\,0}\,t \\
p_{zz}(t)=p_{zz}^{\,0}+2\frac{d}\hbar\,p_{yz}\left(E+\beta B\right)+\frac{\chi}{3}\left(2-p_{zz}^{\,0}\right)\,t-\frac{\chi}{3}\,p_{zz}^{\,0\,2}\,t\nonumber \\
p_{xy}(t)=p_{xy}^{\,0}-\frac{d}\hbar\,p_{xz}\left(E+\beta B\right)+\frac{\chi}{3}\,p_{xy}^{\,0}\,t-\frac{\chi}{3}\,p_{zz}^{\,0}\,p_{xy}^{\,0}\,t\nonumber \\
p_{xz}(t)=p_{xz}^{\,0}+\frac{d}\hbar\,p_{xy}\left(E+\beta
B\right)-\frac{\chi}{6}\,p_{xz}^{\,0}\,t-\frac{3\eta}{4}\,p_{y}^{\,0}\,t
-\frac{\chi}{3}\,p_{zz}^{\,0}\,p_{xz}^{\,0}\,t\nonumber \\
p_{yz}(t)=p_{yz}^{\,0}+\frac{d}\hbar\,(p_{yy}-p_{zz})\left(E+\beta
B\right)-\frac{\chi}{6}\,p_{yz}^{\,0}\,t+\frac{3\eta}{4}\,p_{x}^{\,0}\,t
-\frac{\chi}{3}\,p_{zz}^{\,0}\,p_{yz}^{\,0}\,t. \nonumber
%\eqno(2.28)
\label{2.28}
\end{eqnarray}

According to (\ref{2.28})
 the change in components of vector and tensor polarization caused by the EDM is mixed with the
 contributions from the birefringence effect to the same
 components. In much the same mixing of contributions from the EDM
 and the tensor electric polarizability $\alpha_T$ appear, but
here we does not dwell on them because of the expressions
unhandiness.

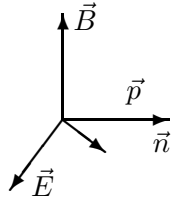
\begin{figure}[h]
\begin{center}
\begin{picture}(40,40)\thicklines
 \put(0,0){\vector(1,0){40}} \put(0,0){\vector(0,1){40}}
 \put(0,0){\vector(-3,-4){20}} \put(34,-12){$\vec{n}$}
 \put(4,34){$\vec{B}$} \put(-12,-29){$\vec{E}$}
 \put(0,0){\vector(4,-3){16}}
 \put(24,8){$\vec{p}$}
\end{picture}
\end{center}
\vspace{1 cm} \caption{The respective orientation of $\vec{E}$,
$\vec{B}$, and $\vec{n}$} \label{Fig.4}
\end{figure}

Thus, the changes in the deuteron  vector
 and tensor polarization are the result of several mechanisms:

- rotation of the spin in the horizontal plane $(\vec{E},\vec{n})$
(Fig.\ref{Fig.4}).

%%%%% place for fig ---- to change the figure !!!!!!!!!!!!!!!!!!!!!!!!!!!!!!!!!!!
%% rotate p in the EN plane !!!!!!!!!!!!!

This rotation is risen due to interaction of the magnetic moment
with external fields. The rotation frequency is expressed as:
\begin{equation}
\vec{\omega}_{a}=\frac{e}{mc}\left\{a\vec{B}+\left(\frac{1}{\gamma^{2}-1}
-a\right)\vec{\beta}\times\vec{E}\right\};
%\eqno(2.29);
\label{2.29}
\end{equation}

\begin{figure}[ht]
\begin{center}
\begin{picture}(40,40)\thicklines
\put(0,0){\vector(1,0){40}}
 \put(0,0){\vector(0,1){40}}
 \put(0,0){\vector(-3,-4){20}}
 \put(34,-12){$\vec{n}$}
 \put(4,34){$\vec{B}$}
 \put(-14,-33){$\vec{E}$}
 \put(0,0){\vector(4,-3){16}}
 \put(24,8){$\vec{p}$}
 \put(0,0){\circle{100}}
\end{picture}
\end{center}
\vspace{1 cm}
\caption{Rotation of the polarization vector due to
birefringence in an electric field} \label{Fig.5}
\end{figure}
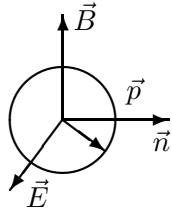

- rotation of the spin in the vertical plane $(\vec{B},\vec{n})$
caused by the electric dipole moment (Fig.\ref{Fig.5});

% place for fig
%%% the place for figure change the figure !!!!!!!!!!!!!!!!!!!!!!!!!!!!!!!!!!1

The rotation frequency is
\begin{equation}
\vec{\omega}_{d}=d\frac{c}{\hbar}\left(\frac{\vec{E}}{c}+\vec{\beta}\times\vec{B}\right)
%\eqno(2.30);
\label{2.30}
\end{equation}

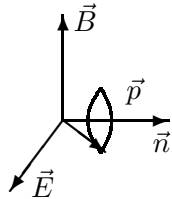
\begin{figure}[ht]
\begin{center}
\begin{picture}(40,40)\thicklines
\put(0,0){\vector(1,0){40}}
 \put(0,0){\vector(0,1){40}}
 \put(0,0){\vector(-3,-4){20}}
 \put(34,-12){$\vec{n}$}
 \put(4,34){$\vec{B}$}
 \put(-12,-29){$\vec{E}$}
 \put(0,0){\vector(4,-3){16}}
 \put(24,8){$\vec{p}$} \
 \qbezier(14,-12)(4,0)(14,12)
 \qbezier(14,-12)(22,0)(14,12)
\end{picture}
\end{center}
\vspace{1 cm} \caption{Rotation of the polarization vector due to
the phenomena of birefringence} \label{Fig.6}
\end{figure}

- rotation caused by the phenomenon of birefringence in a medium,
this is precession in the vertical plane $(\vec{B},\vec{E})$
(Fig.\ref{Fig.6});

%%% place for figure

The rotation frequency is
\begin{equation}
\omega=\frac{2\pi N}{M
\gamma}\,\hbar\,\texttt{Re}d_{1}.
%\eqno(2.31)
\label{2.31}
\end{equation}

\begin{figure}[ht]
\begin{center}
\begin{picture}(40,40)\thicklines
\put(0,0){\vector(1,0){40}} \put(0,0){\vector(0,1){40}}
\put(0,0){\vector(-3,-4){20}} \put(34,-10){$\vec{n}$}
\put(4,34){$\vec{B}$} \put(-14,-28){$\vec{E}$}
\put(0,0){\vector(4,-3){16}} \put(24,8){$\vec{p}$}
\put(0,0){\circle{100}}
\end{picture}
\end{center}
\caption{Rotation of the polarization vector due to birefringence
in an electric field} \label{Fig.5b}
\end{figure}
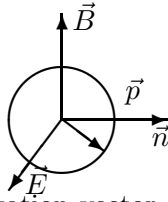

- rotation due to the phenomenon of birefringence in an electric
field in the vertical plane $(\vec{B},\vec{n})$
(Fig.\ref{Fig.5b}), i.e. in the same plane as the rotation caused
by the EDM.

% place for fig
%%% the place for figure change the figure !!!!!!!!!!!!!!!!!!!!!!!!!!!!!!!!!!1

The rotation frequency is
\begin{equation}
\omega_{E}=\frac{\alpha_{T}E_{eff}^{2}}{\hbar}
%\eqno(2.32)
\label{2.32}
\end{equation}

Besides the above rotations \textbf{the transitions from the
vector polarization into the tensor one and spin dichroism occur.}

Moreover, the spin dichroism leads to the appearance of the tensor
polarization.

 Let us compare the frequency and the angle of polarization vector
rotation caused by the EDM with those caused by the birefringence
effect.

1. The spin rotation frequency caused by the EDM is determined by
the formula (\ref{2.3})
\begin{equation}
\omega_{edm}=\frac{d\,E}{\hbar}+\frac{d}{\hbar}\beta B.
%\eqno(2.33)
\label{2.33}
\end{equation}
We can get  $\omega_{edm}\approx 3\cdot10^{-7}rad/s$ for the
storage ring with $E=3.5\,MV/m$, $B=0.2\,T$ and expected value of
the deuteron EDM $d\sim10^{-27}e\cdot cm$ and $\omega_{edm}\approx
3\cdot10^{-9}rad/s$ for EDM $d\sim10^{-29}e\cdot cm$.

2. The spin rotation frequency caused by the phenomena of
birefringence in a residual gas:
\begin{equation}
\omega=\frac{2\pi N}{M \gamma}\,\hbar \,\texttt{Re}d_{1}.
%\eqno(2.34)
\label{2.34}
\end{equation}
Using the last formula one can get
$\omega\approx2\cdot10^{-7}\,rad/s$ for $N=10^{9}\,cm^{-3}$
(suppose the pressure inside the storage ring $\sim 10^{-7}$
Torr), $\texttt{Re}d_{1}\sim10^{-13}$. This effect depends on the
density $N$ (depends on the pressure inside the storage ring).

3. The spin rotation frequency caused by the phenomenon of
birefringence in the gas jet (gas target), which is used for the
beam extraction to the polarimeter \cite{project}
(Fig.\ref{Polarimeter}):

\begin{figure}
\epsfxsize = 8 cm \centerline{\epsfbox{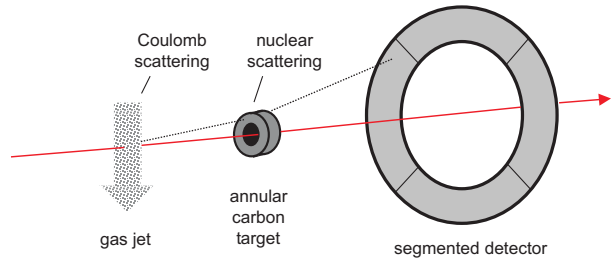}}
\caption{Polarimeter \cite{project}} \label{Polarimeter}
\end{figure}
 In this case  the effective rotation  frequency (or the rotation angle
for $\tau=1s$) is determined as follows:
\begin{equation}
\omega_{t\,eff}\equiv\varphi_{t}=\frac{2\pi
N_{t}}{k}\,l\,\texttt{Re}d_{1}\,\nu,
%\eqno(2.35)
\label{2.35}
\end{equation}
where  $\omega_{t\,eff}$ is the effective frequency of spin
rotation , $\varphi_{t}$ is the deuteron spin rotation angle for
$\tau=1s$,  $N_{t}$ is the density of the target, $l$ is the
length of the target, $\nu$ is the frequency of beam rotation in
the storage ring, $k=\frac{M \gamma v}{\hbar}$ is the deuteron
wave number.

 Really, the frequency $\omega_{t}$ of the spin rotation in
 matter is
\begin{equation}
\omega_{t}=\frac{2\pi N_{t}}{M
\gamma}\,\hbar\,\texttt{Re}d_{1}
%\eqno(2.36)
\label{2.36}
\end{equation}
The spin rotation angle $\theta_{\tau}=\omega_t \tau_t$, where
$\tau_t=\frac{l}{v}$ is the deuteron flying time in the target.
The angle of rotation at 1 second is $\theta=\omega_t \tau_t \nu$
(during a second the deuteron passes through the gas target $\nu$
times). As a result,
\begin{equation}
\omega_{t\,eff}=\omega_{t}\,\tau_{t}\,\nu,
%\eqno(2.37)
\label{2.37}
\end{equation}
so one can get
\begin{equation}
\omega_{t\,eff}=\frac{2\pi N_t}{M
\gamma}\,\hbar\,\texttt{Re}d_{1}\frac{l}{v}\,\nu=\frac{2\pi
j}{k}\,\texttt{Re}d_{1} \nu,
%\eqno(2.38)
\label{2.38}
\end{equation}
where $j=N_t\,l$. Then using the experimental parameters
\cite{project} $j=10^{15}\,cm^{-2}$, $\nu\approx 10^{5}-10^{6}$ we
have $\omega_{t\,eff}\approx 10^{-5}-10^{-4}\,rad/s$.

So the angle of polarization vector rotation  for $\tau=1s$ caused
by the phenomenon of birefringence in the gas jet of polarimeter
appears by two orders of magnitude greater than the angle of
rotation due to the EDM. The additional contribution to spin
rotation is also provided by the solid carbon target of
polarimeter (see Figure.\ref{Polarimeter}).

4. The estimation for the spin rotation frequency caused by the
birefringence in an electric field according to (\ref{2.30}) for
$\alpha_{T}\sim10^{-37}cm^{3}$ and $E=3.5MV/m$ is
$\omega_{E}\sim10^{-6}rad/s$.

Let us estimate the value of spin dichroism. This characteristic
is given by the parameter $\chi=-vN(\sigma_1-\sigma_0)$ (see
expressions (\ref{2.25}), (\ref{2.26})):

- in the case of the scattering by the residual gas we have
$\left| \chi \right| \sim 0.5\cdot10^{-6}$ for $N \sim 10^{9}$
cm$^{-3}$;

- if a deuteron passes through the gas jet (gas target) for
$\tau=1$s is
\begin{equation}
\chi_{t}={j}(\sigma_{1}-\sigma_{0})\,\nu,
%\eqno(2.40),
\label{2.40}
\end{equation}
then for $j=10^{15}\,cm^{-2}$ and $\nu\approx6\cdot10^{5}$ one can
get $\chi_{t}\sim 3.4\cdot10^{-5}$. So we can conclude that there
is the significant beam spin dichroism.

It should be especially mentioned that the polarimeter
(Fig.\ref{Polarimeter}) includes a solid target. Therefore, the
birefringence effect in the matter of the solid target arises,
too.

The possible influence of P-odd spin rotation and spin dichroism
(i.e. different absorbtion cross-sections in matter for a deuteron
with the spin parallel and antiparallel to the momentum
$\sigma_{\uparrow \uparrow} \ne \sigma_{\uparrow \downarrow}$)
aren't to be forgotten.

\section{CONCLUSION}

 The above analysis shows that the phenomena of deuteron
 birefringence in matter and an electric field should be
 studied in experiments for the EDM
 search with a storage ring.
Moreover, they could imitate the spin
 rotation due to the EDM.
Study of the birefringence effects in such
 experiments could
 provide to measure both the spin-dependent part $d_1$ of the amplitude of
 the coherent elastic scattering of a deuteron by a nucleus at
 the zero angle and the tensor electric polarizability of a deuteron.

 It should be also mentioned that if the nuclei in the gas jet are
 polarized, then according to \cite{4} the P-,T-odd spin rotation and
 dichroism appear in the storage ring. They are caused by the T-odd
 nucleon-nucleon
 interaction of a deuteron with a polarized nucleus and, in particular, interaction
 described by $V_{P,T} \sim \vec{S} \left[ \vec{p}_N \times \vec{n}
 \right]$, where $\vec{p}_N$ is the polarization vector of gas
 target.

  P-even T-odd spin rotation and dichroism of deuterons (nuclei)
  caused by the interaction either $V_T \sim (\vec{S} [ \vec{p}_N \times \vec{n}
  ])(\vec{S}\vec{n})$ or
  $V^{\prime}_T \sim Sp \rho_J [([\vec{S} \times \vec{n}] \vec{J}) (\vec{J} \vec{n})]$
  also could be observed
  \cite{4} (here $J \ge 1$ is the spin of the polarized target
  nuclei, $\rho_J$ is the spin matrix density of the target
  nuclei).

\section{Appendix}
Let us consider the amplitude of deuteron scattering by a proton
depending on spin orientation with respect to the deuteron
momentum \cite{2}.

For fast deuterons the scattering amplitude can be found in the
eikonal approximation \cite{Cryz,hand}. According to \cite{hand},
the amplitude of zero-angle coherent scattering in this
approximation can be written as follows:
%\bibitem{Cryz} W. Cryz, L.C. Maximon Ann.Phys., NY {\bf 52} (1969) 59
%\bibitem{hand} Handbuch der Physik {\bf 39} (Berlin: Springer, 1957) 112
\begin{eqnarray}
f(0)=\frac{k}{2\pi~i}\int \left( e^{i\chi _{D}\left( \overrightarrow{b},\overrightarrow{r}%
\right) }-1\right) d^{2}b\left| \varphi \left(
\overrightarrow{r}\right) \right| ^{2}d^{3}r \label{amp}
\end{eqnarray}
where $k$ is the deuteron wavenumber, $\overrightarrow{b}$ is the
impact parameter, i.e. the distance between the deuteron and the
proton centers of gravity in the normal plane; $\varphi \left(
\overrightarrow{r}\right)$ is the wavefunction of the deuteron in
the ground state; $\left| \varphi \left(
\overrightarrow{r}\right)\right| ^{2}$ is the probability to find
proton and neutron (in the deuteron) at the distance
$\overrightarrow{r}$ apart. The phase shift due to the deuteron
scattering by a proton is
\begin{equation}
\chi _{D}=-\frac{1}{\hbar v}
\int_{-\infty }^{+\infty }V_{D}\left( \overrightarrow{b},z^{^{\prime }},%
\overrightarrow{r}_{\perp }\right) dz^{^{\prime }}
\end{equation}
$\overrightarrow{r}_{\perp}$ is the $\overrightarrow{r}$ component
perpendicular to the momentum of the incident deuteron, $v$ is the
deuteron velocity. The phase shift $\chi _{D}=\chi _{1}+\chi
_{2}$, where $\chi_{1}$ and $\chi _{2}$ are the phase shifts
caused by proton-proton and neutron-proton interactions,
respectively.

For the polarized deuteron under consideration the probability
$\left| \varphi \left( \overrightarrow{r}\right)\right| ^{2}$ is
different for the different spin states of the deuteron. Thus for
states with the magnetic quantum number $m=\pm 1$, the probability
is $\left| \varphi_{\pm 1} \left( \overrightarrow{r}\right)\right|
^{2}$, whereas for $m=0$, it is $\left| \varphi_{0} \left(
\overrightarrow{r}\right)\right| ^{2}$. Owing to the additivity of
the phase shifts, the equation (\ref{amp}) can be rewritten as
\begin{equation}
f\left( 0\right) =\frac{k}{\pi }\int \left\{
t_{1} \left(
\overrightarrow{b}-%
\frac{\overrightarrow{r}_{\perp }}{2} \right) +t_{2} \left(
\overrightarrow{b}+%
\frac{\overrightarrow{r}_{\perp }}{2} \right) + 2it_{1} \left(
\overrightarrow{b }-\frac{\overrightarrow{r}_{\perp }}{2} \right)
t_{2} \left(
\overrightarrow{b}%
+\frac{\overrightarrow{r}_{\perp }}{2} \right)
\right\} \left| \varphi \left( \overrightarrow{r} \right) \right|
^{2} d^{2}bd^{3}r \label{42}
\end{equation}
where
\[
t_{1(2)}=\frac{e^{i\chi _{1\left( 2\right) }}-1}{2i}.
\]

Attention should be given to the fact that the latter expression
is valid if one neglects spin dependence of nuclear forces between
the colliding proton (neutron) and proton. When this dependence is
taken into account, the phase shift $\chi_{D}$ is an operator
acting in the spin space of colliding particles and in the general
case the expansion (\ref{42}) is not valid. However, to estimate
the magnitude of the effect of deuteron spin oscillations, spin
dependence of nuclear forces may be neglected (spin dependent
contribution was considered in \cite{kostya}). Let us also omit
terms caused by the Coulomb interaction. From (\ref{42}) it
follows
%\onecolumn
\begin{equation}
f(0)=f_{1}(0)+f_{2}(0)+ \frac{2ik}{\pi}
\int t_{1}\left( \overrightarrow{b}-%
\frac{\overrightarrow{r}_{\perp }}{2}\right)t_{2}\left( \overrightarrow{b}+%
\frac{\overrightarrow{r}_{\perp }}{2}\right)\left| \varphi \left(
\overrightarrow{r}_{\perp},z\right) \right|
^{2}d^{2}bd^{2}r_{\perp}dz \label{integral}
\end{equation}
where
\[
f_{1(2)}(0)=\frac{k}{\pi} \int
t_{1(2)}(\overrightarrow{\xi})d^{2}\xi=
\frac{m_D}{m_{1(2)}}~f_{p(n)}(0)
\]
and $f_{p(n)}(0)$ is the amplitude of the proton (neutron)-proton
zero-angle elastic coherent scattering.

The expression (\ref{integral}) can be rewritten as
\begin{equation}
f(0)=f_{1}(0)+f_{2}(0)+ \frac{2ik}{\pi}\int
t_{1}(\overrightarrow{\xi})~ t_{2}(\overrightarrow{\eta}) \left|
\varphi \left(\overrightarrow{\xi}-\overrightarrow{\eta},z\right)
\right| ^{2}~d^{2}\xi~d^{2}\eta~dz \label{27}
\end{equation}

Then from (\ref{27})
\begin{eqnarray}
&\Re& f(0)=\Re~f_{1}(0)+\Re~f_{2}(0) -\frac{2k}{\pi} \Im \int
t_1(\overrightarrow{\xi}) t_{2}(\overrightarrow{\eta})\left|
\varphi \left(\overrightarrow{\xi}-\overrightarrow{\eta},z\right)
\right| ^{2}~d^{2}\xi~d^{2}\eta~dz \\ \nonumber &\Im&
f(0)=\Im~f_{1}(0)+\Im~f_{2}(0)+ +\frac{2k}{\pi}\Re \int
t_1(\overrightarrow{\xi}) t_{2}(\overrightarrow{\eta})\left|
\varphi \left(\overrightarrow{\xi}-\overrightarrow{\eta},z\right)
\right| ^{2}~d^{2}\xi~d^{2}\eta~dz \nonumber \label{28}
\end{eqnarray}

%%%%%%%%%%%%!!!!!!!!!!!!!!!!!!!!!!!!!!!!!!!!!!!!!!!!!!!!!!!!!!!!!!!!!!!!!!!!!!!!11

The spin oscillation period is determined by the difference in the
amplitudes $\Re~f(m=\pm1)$ and $\Re~f(m=0)$. From (\ref{42}) it
follows that \
\begin{small}
\begin{eqnarray}
\Re~d_1=-\frac{2k}{\pi} \Im \int t_1(\overrightarrow{\xi})
t_{2}(\overrightarrow{\eta})\left[ \varphi_{\pm 1}^{+}
\left(\overrightarrow{\xi}-\overrightarrow{\eta},z\right)
\varphi_{\pm 1}
\left(\overrightarrow{\xi}-\overrightarrow{\eta},z\right)-
\varphi_{0}^{+}
\left(\overrightarrow{\xi}-\overrightarrow{\eta},z\right)
\varphi_{0}
\left(\overrightarrow{\xi}-\overrightarrow{\eta},z\right)
\right]~d^{2}\xi~d^{2}\eta~dz \nonumber \\
\Im~d_1=\frac{2k}{\pi}\Re \int t_1(\overrightarrow{\xi})
t_{2}(\overrightarrow{\eta})\left[ \varphi_{\pm 1}^{+}
\left(\overrightarrow{\xi}-\overrightarrow{\eta},z\right)
\varphi_{\pm 1}
\left(\overrightarrow{\xi}-\overrightarrow{\eta},z\right)-
\varphi_{0}^{+}
\left(\overrightarrow{\xi}-\overrightarrow{\eta},z\right)
\varphi_{0}
\left(\overrightarrow{\xi}-\overrightarrow{\eta},z\right)
\right]~d^{2}\xi~d^{2}\eta~dz \nonumber \label{d1}
\end{eqnarray}
\end{small}
It is well known that the characteristic radius of the deuteron is
large comparing with the range of nuclear forces. For this reason,
when integrating, the functions $t_1$ and $t_2$ act on $\varphi$
as $\delta$-function. Then

\begin{eqnarray}
\Re~d_1=-\frac{4k}{\pi}\Im {f_1(0)~f_{2}(0)} \int_{0}^{\infty}
\left[ \varphi_{\pm 1}^{+} \left(0,z\right) \varphi_{\pm 1}
\left(0,z\right)- \varphi_{0}^{+} \left(0,z\right) \varphi_{0}
\left(0,z\right) \right]~dz \\ \nonumber \Im~d_1=\frac{4k}{\pi}\Re
{f_1(0)~f_{2}(0)} \int_{0}^{\infty} \left[ \varphi_{\pm 1}^{+}
\left(0,z\right) \varphi_{\pm 1} \left(0,z\right)- \varphi_{0}^{+}
\left(0,z\right) \varphi_{0} \left(0,z\right) \right]~dz
\label{d1}
\end{eqnarray}

The magnitude of the spin oscillation effect is determined by the
difference
\[
\left[ \varphi_{\pm 1}^{+} \left(0,z\right) \varphi_{\pm 1}
\left(0,z\right)- \varphi_{0}^{+} \left(0,z\right) \varphi_{0}
\left(0,z\right)\right]
\]
i.e. by the difference in distributions of nucleon density in the
deuteron for different deuteron spin orientations. The structure
of the wavefunction $\varphi_{\pm 1}$ is well known:
\begin{equation}
\varphi_m=\frac{1}{4 \pi} \left\{ \frac{u(r)}{r}+\frac{1}{\sqrt
8}\frac{W(r)}{r}\hat{S}_{12} \right\} \chi_m \label{phi_m}
\end{equation}
where $u(r)$ is the deuteron radial wavefunction corresponding to
the S-wave; $W(r)$ is the radial function corresponding to the
D-wave; the operator $\hat{S}_{12}=6(\hat{\overrightarrow{S}}
\overrightarrow{n}_{r})^2-2\hat{\overrightarrow{S}}^2$;
$\overrightarrow{n}_{r}=\frac{\overrightarrow{r}}{r}$;
$\hat{\overrightarrow{S}}=\frac{1}{2}(\overrightarrow{\sigma}_1+\overrightarrow{\sigma}_2)$
and $\overrightarrow{\sigma}_{1(2)}$ ate the Pauli spin matrices
describing proton(neutron) spin.

Use of (\ref{phi_m}) yields
\begin{eqnarray}
\Re~d_1=-\frac{3}{2 \pi k}~\Im \left\{ f_{1}(0)f_{2}(0) \right\}
\textrm{D} = -\frac{\sigma}{\pi k} \Im \left\{ f_{p}(0)f_{n}(0)
\right\} \textrm{D}
\\ \nonumber \Im~d_1=\frac{3}{2 \pi k}~\Re \left\{ f_{1}(0)f_{2}(0)
\right\} \textrm{D}= \frac{\sigma}{\pi k} \Re \left\{
f_{p}(0)f_{n}(0) \right\} \textrm{D}
\end{eqnarray}
where $\textrm{D}=\int_{0}^{\infty} \left( \frac{1}{\sqrt
2}\frac{u(r)W(r)}{r^2}-\frac{1}{4} \frac{W^2(r)}{r^2} \right) dr
$, $r^2=\xi^2+z^2$. Applying the optical theorem $\Im~f=\frac{k}{4
\pi}~\sigma$, where $\sigma$ is the total scattering
cross-section, one can obtain
\begin{eqnarray}
\Re~d_1=-\frac{3}{2 \pi^2} \left( \Re~f_p(0) \sigma_{n}+\Re~f_n(0)
\sigma_{p}
\right) \textrm{D} \\
\Im~d_1=\frac{3}{2 \pi k} \left( \Re~f_1~\Re~f_2-\frac{k^2}{(4
\pi)^2}\sigma_{1}\sigma_{2} \right) \textrm{D} = \left(
\frac{6}{\pi k} \Re~f_p~\Re~f_n-\frac{3k}{(2
\pi)^3}\sigma_{p}\sigma_{n} \right) \textrm{D}
\end{eqnarray}
where $\sigma_{p(n)}$ is the total cross-section of the
proton-proton (neutron-proton) nuclear scattering.

\begin{figure}[htbp]
\begin{minipage}[htbp]{3.00 in}
\epsfxsize=7.0 cm \epsfbox{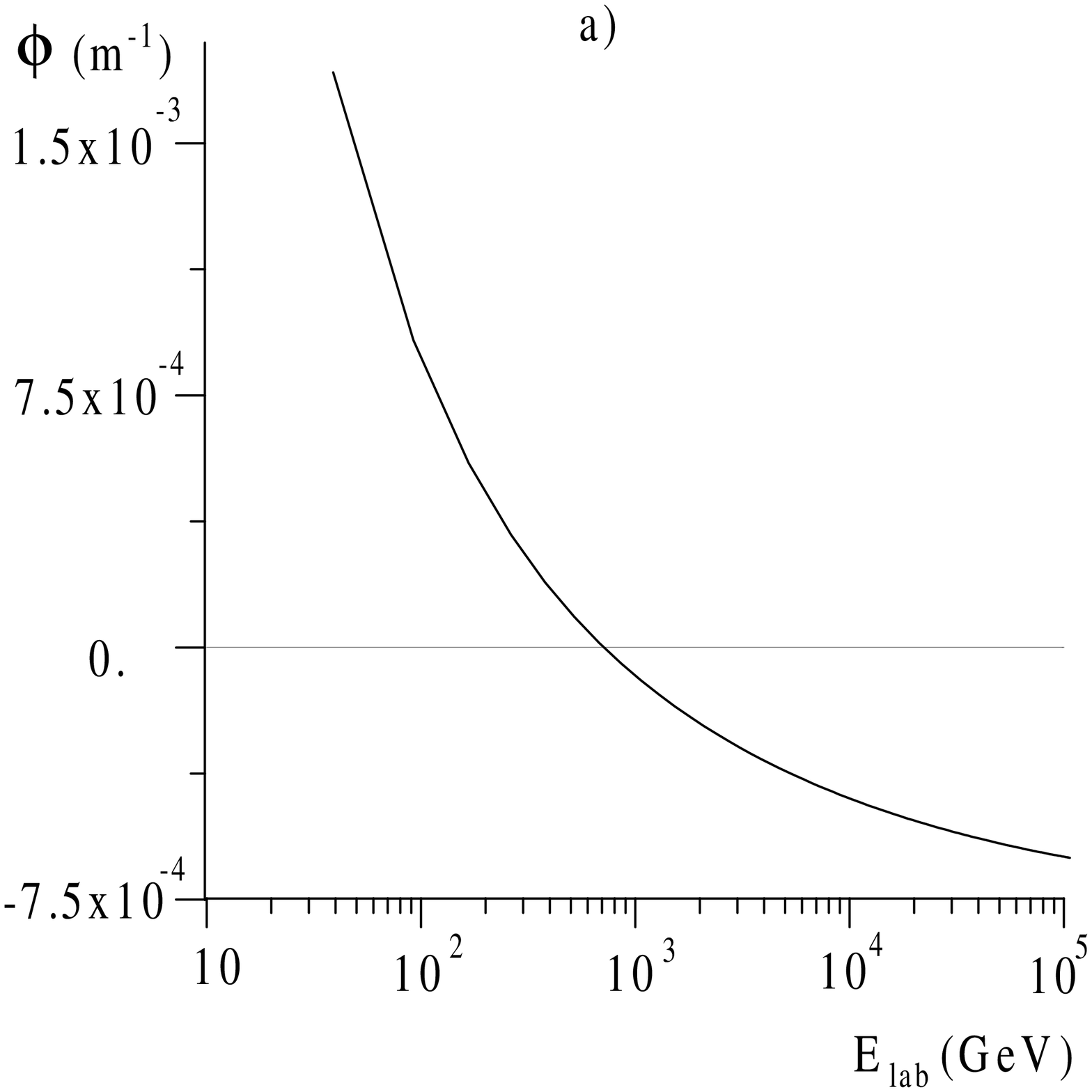}
\end{minipage}
\begin{minipage}[htbp]{3.00 in}
\epsfxsize=7.0 cm \epsfbox{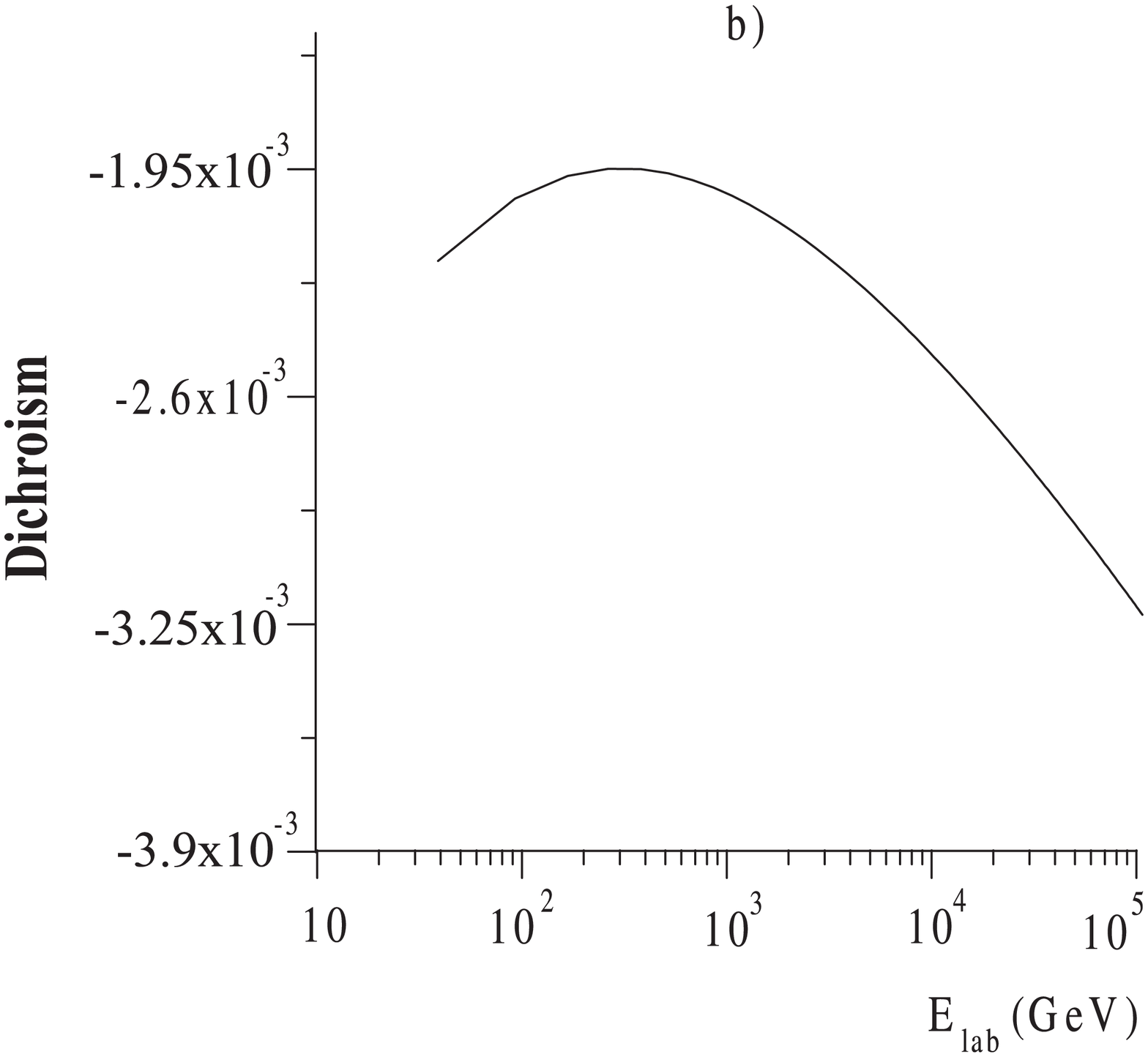}
\end{minipage}
\caption{Rotation angle and dichroism in hydrogen target at high
energies} \label{high}
\end{figure}

\begin{figure}[htbp]
\begin{minipage}[htbp]{3.00 in}
\epsfxsize=7.0 cm \epsfbox{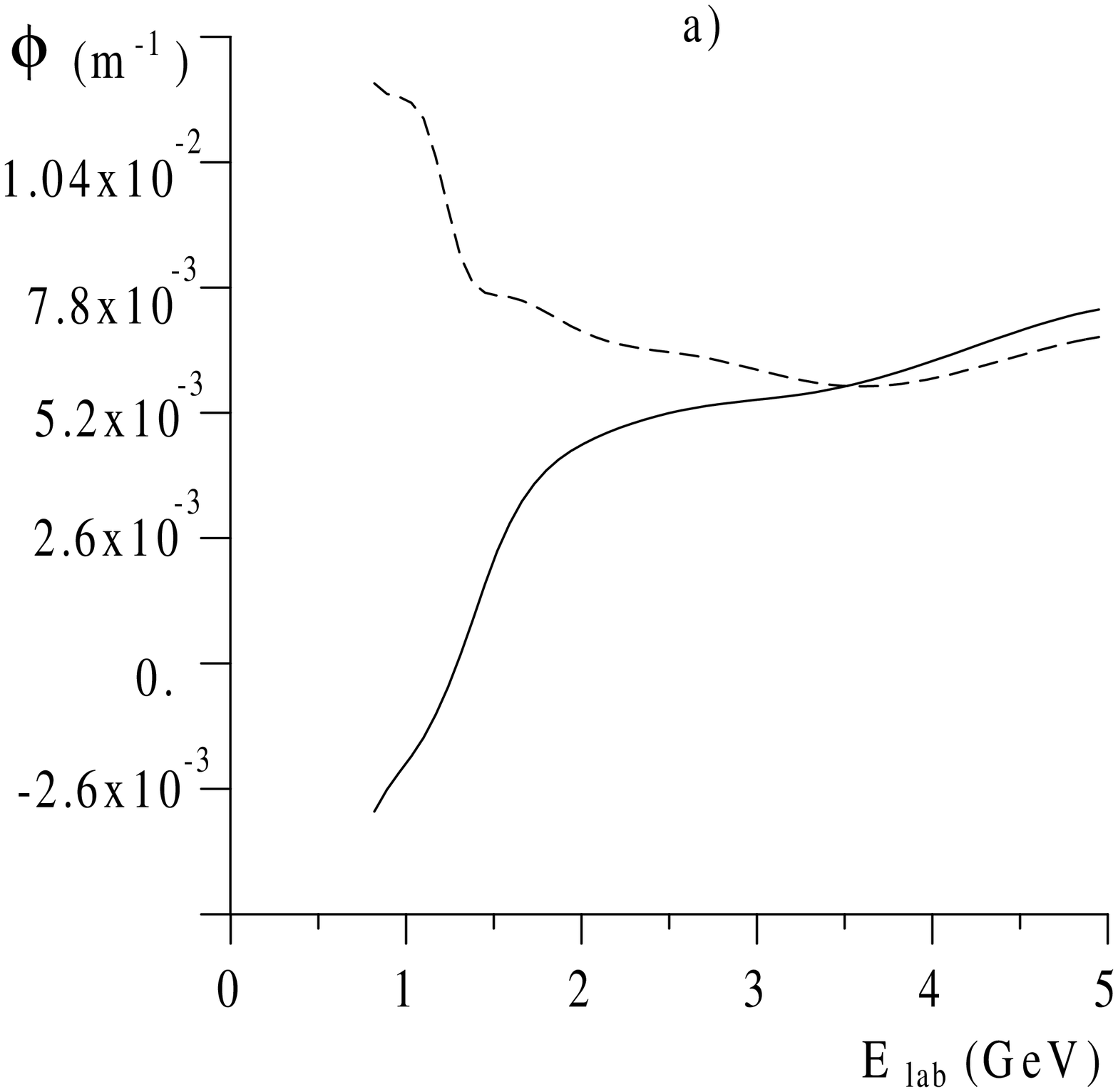}
\end{minipage}
\begin{minipage}[htbp]{3.00 in}
\epsfxsize=7.0 cm \epsfbox{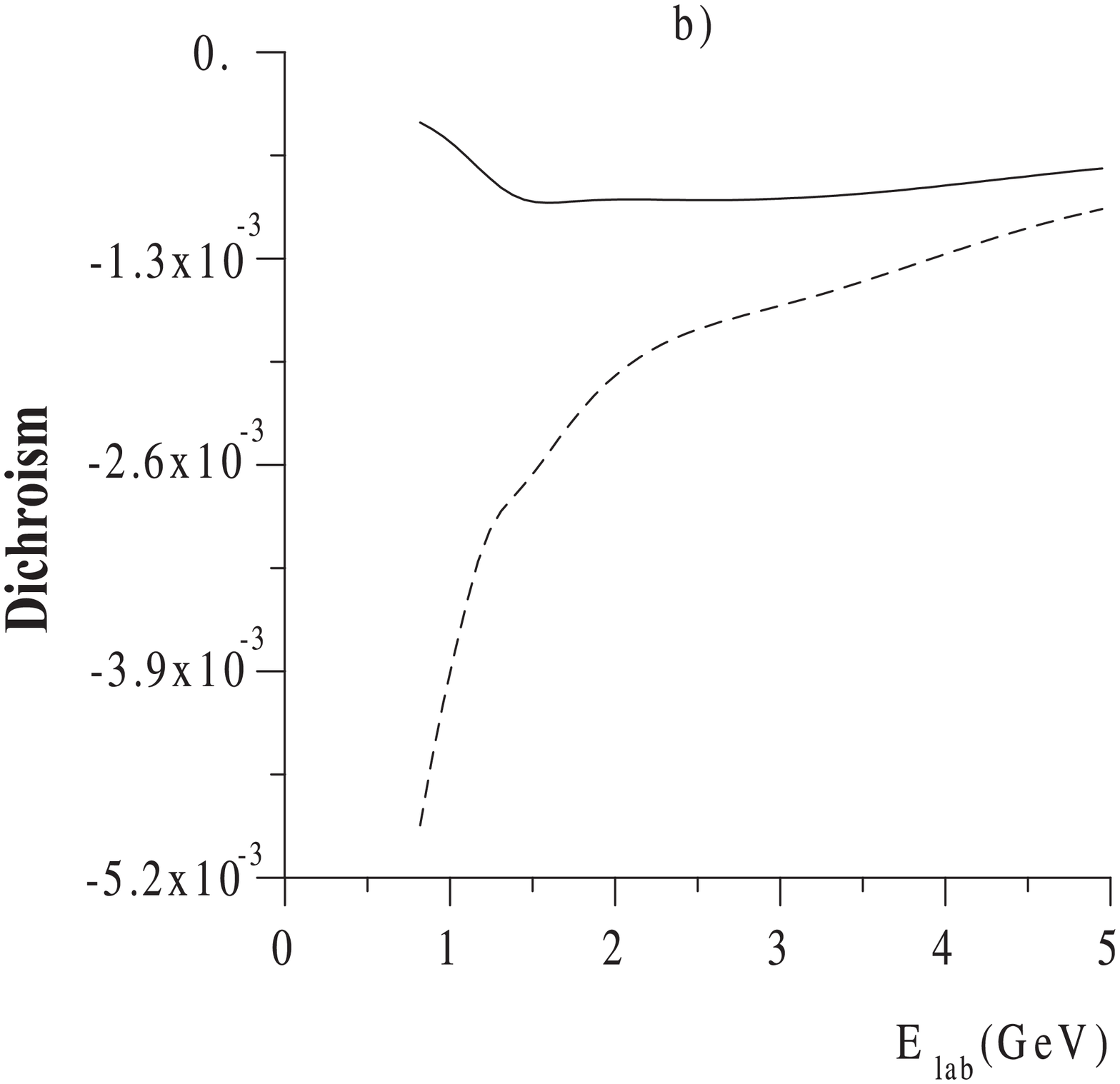}
\end{minipage}
\caption{Spin rotation angle and dichroism in hydrogen target at
deuteron energies lower than 5 Gev. Solid curve corresponds to the
calculation with the spinless $N-N$ amplitudes} \label{low}
\end{figure}

% ---------------------------------- from wind - literature correction

\end{document}